\documentclass{aa}  

\usepackage[normalem]{ulem}
\usepackage{graphicx}
\usepackage{color}
\usepackage{url}
\usepackage{hyperref}
\usepackage{lipsum}
\usepackage{multirow}
\usepackage{xcolor}

\usepackage{txfonts}
\usepackage{natbib}
\usepackage{amsmath}    
\usepackage{amssymb}    
\newcommand{\cyg}{Cyg~X\mbox{-}1\xspace}
\newcommand{\DUF}{DIPol-UF\xspace}
\newcommand{\DP}{DIPol-2\xspace}

\newcommand{\unit}[1]{\mbox{\boldmath $\hat{#1}$}}
\newcommand{\rchi}{\chi}

\newcommand{\qobs}{$q_\text{obs}$\xspace}
\newcommand{\uobs}{$u_\text{obs}$\xspace}

\newcommand{\qint}{$q_\text{int}$\xspace}
\newcommand{\uint}{$u_\text{int}$\xspace}

\newcommand{\qis}{$q_\text{IS}$\xspace}
\newcommand{\uis}{$u_\text{IS}$\xspace}

\DeclareMathOperator{\arctantwo}{arctan2}

\begin{document} 


\title{Peering into the tilted heart of \cyg with high-precision optical polarimetry} 

\titlerunning{Optical polarization of \cyg }

\author{Vadim Kravtsov\inst{1}
\and  
Alexandra Veledina\inst{1,2}   
\and 
Andrei V. Berdyugin\inst{1}
\and
Andrzej A. Zdziarski\inst{3}
\and
Gary D. Henson\inst{4}
\and
Vilppu Piirola\inst{1}   
\and 
Takeshi Sakanoi\inst{5}   
\and 
Masato Kagitani\inst{5}   
\and 
Svetlana V. Berdyugina\inst{6,7}   
\and
Juri Poutanen\inst{1}}

\authorrunning{V. Kravtsov et al.} 
 
\institute{Department of Physics and Astronomy, FI-20014 University of Turku, Finland\\
\email{vakrau@utu.fi}
\and
Nordita, KTH Royal Institute of Technology and Stockholm University, Hannes Alfv\'ens v\"ag 12, SE-10691 Stockholm, Sweden 
\and 
Nicolaus Copernicus Astronomical Center, Polish Academy of Sciences, Bartycka 18, PL-00-716 Warszawa, Poland
\and
Department of Physics and Astronomy, East Tennessee State University, Johnson City, TN 37614, USA
\and
Graduate School of Sciences, Tohoku University, Aoba-ku,  980-8578 Sendai, Japan
\and
Leibniz-Institut f\"{u}r Sonnenphysik, Sch\"{o}neckstr. 6, 79104 Freiburg, Germany
\and
Istituto Ricerche Solari Aldo e Cele Dacc\`o (IRSOL), Faculty of Informatics, Universit\`a della Svizzera italiana, 6605 Locarno, Switzerland 
}


\abstract{
We present the high-precision optical polarimetric observations of black hole X-ray binary \cyg, spanning several cycles of its 5.6~day orbital period.
Week-long observations on two telescopes located in opposite hemispheres allowed us to track the evolution of the polarization within one orbital cycle with the highest temporal resolution to date. 
Using the field stars, we determine the interstellar polarization in the source direction and subsequently its intrinsic polarization.
The optical polarization angle is aligned with that in the X-rays as recently obtained with the Imaging X-ray Polarimetry Explorer. 
Furthermore, it is consistent, within the uncertainties, with the position angle of the radio ejections.
We show that the intrinsic PD is variable with the orbital period with the amplitude of $\sim$0.2\% and discuss various sites of its production.
Assuming the polarization arises from a single Thomson scattering of the primary star radiation by the matter that follows the black hole in its orbital motion, we constrain the inclination of the binary orbit $i>120\degr$ and its eccentricity $e<0.08$.
The asymmetric shape of the orbital profiles of Stokes parameters implies also the asymmetry of the scattering matter distribution about the orbital plane, which may arise from the tilted accretion disk.
We compare our data to the polarimetric observations made over 1975--1987 and find good, within 1\degr, agreement between the intrinsic polarization angles.
On the other hand, the PD decreased by 0.4\% over half a century, suggesting the presence of secular changes in the geometry of accreting matter.
}


\maketitle 
%

\section{Introduction}

Determination of the large-scale accretion geometry and orbital parameters is a problem of fundamental importance for X-ray binaries.
Various techniques can be employed to examine the geometry of these systems, e.g. photometry, spectroscopy, imaging, and timing, but a special place in this list belongs to polarimetry, which is known to be most sensitive to changes in geometry. 
The geometrical properties can be determined by tracking the changes in polarization degree (PD) and polarization angle (PA) as a function of the orbital phase.
The stochastic variability at timescales comparable to the orbital period may significantly alter the average polarization profile.
Dense coverage of an entire orbital cycle is needed to reliably determine the accretion geometry, shape, and orientation of the binary components. 

The orbital parameters in binary systems are conventionally studied using optical and infrared polarimetry.
For the low-mass X-ray binaries in outburst, emission in these wavelengths can be composed of several components: an (irradiated) accretion disk, wind, jet, and hot accretion flow \citep{PV2014, UC2014}.
Optical polarimetry has been used as a fine tool to discriminate between them \citep{Veledina2019, Kosenkov2020b}.
In the (near-)quiescence, the optical polarimetry has helped constrain the role of the non-stellar components in total spectra \citep{Kravtsov2022}, and to determine the misalignment of the black hole (BH) and orbital spins \citep{Poutanen2022}.
For the high-mass X-ray binaries, emission in infrared, optical, and ultraviolet bands is completely dominated by the donor star, which can be scattered by different large-scale components in the binary -- the accretion stream, disk, or outflow/jet. 
Polarization signal in this case can reveal the location, orientation, and physical properties of the scattering component \citep{Jones1994}.

With the launch of the {Imaging X-ray Polarimetry Explorer} \citep[{IXPE},][]{Weisskopf2022}, the polarimetric field gained a new breath. 
It became possible to directly link the orientation of the large-scale binary components, probed by the optical and infrared wavelengths, to the innermost accretion geometry, using X-ray polarimetry.
The prototypical BH X-ray binary Cyg~X-1 became the first target of such studies \citep{Krawczynski2022}.
The week-long IXPE exposure has been accompanied by the global multiwavelength campaigns, allowing it to cover a big fraction of its 5.6~d orbital period.

Cyg X-1 is the first discovered BH X-ray binary and a well-studied system \citep{Bowyer1965}.
It is a persistent source and a high-mass binary hosting a supergiant $\sim 40 M_\odot$ donor star of spectral type O in a nearly-circular orbit (eccentricity $e \sim 0.02$) with the most massive Galactic BH $M_{\rm BH}=21.2\pm2.1 M_\odot$ known to date \citep{Miller-Jones2021}.
The donor is close to filling its Roche lobe  and the compact object accretes the matter through the focused stellar wind \citep{GiesBolton1986b}. 
Accretion proceeds through the disk whose emission is often seen in the  X-rays \citep{Gierlinski1997,Gierlinski1999,ZG2004} and a fraction of matter leaves the system in the form of the jet \citep{Stirling2001, Fender2006, Miller-Jones2021}.

Optical radiation is dominated by the light of the companion star and shows pronounced variations at an orbital period caused by the asymmetric shape of the donor \citep{Kemp1983, GiesBolton1986, Brocksopp1999b, Orosz2011}.
Optical polarization measurements also show pronounced orbital variability \citep{Nolt1975}.
The observed double-peak sinusoidal variations of the PD and PA are consistent with the scenario where polarization arises from Thomson scattering of the donor star radiation by an optically thin matter located within the binary \citep*{BME, Milgrom1978}.
This pattern is typical for binary systems and was observed in a variety of sources from classical \citep{Piirola1980, Berdyugin1999} to gamma-ray binaries \citep{Kravtsov2020}.
The synchronization with the orbital phase indicates that the source of the polarization is connected to the orbital motion of the BH around the companion star.

Polarization may originate from the accretion stream, its impact point in the accretion disk or the disk matter itself, or be related to the outflow/jet.
Which component causes variations of polarization in Cyg X-1 is unknown.
The shapes of the PD and PA profiles have been used in several works to constrain the orbital parameters such as inclination and eccentricity \citep{Kemp1978,Kemp1983,Karitskaya1981,Dolan1989,Nagae2009}.

The long orbital period of Cyg X-1 became an obstacle in tracking the polarimetric variations over a large fraction of a single cycle, and the average profile was obtained by including data from many cycles.
This approach may however lead to a substantial deviation of the obtained mean profile from the individual cycles, as the system is known to show substantial superorbital variability \citep{Priedhorsky1983,Karitskaya2001,Poutanen2008, Zdziarski2011}.
This means that the scattering matter gradually rearranges within the binary over the superorbital period, leading to a systematic bias in the determination of the binary inclination from the mean orbital profile.
On the other hand, the data obtained during a single orbital period at one telescope unavoidably suffers from under-sampling \citep{Dolan1989,Nagae2009}. 
We performed multi-observatory polarimetric observations of Cyg X-1, for the first time covering up to 30\% of the orbit in one cycle.
This was achieved by observing with nearly identical polarimeters from the telescopes separated by $\sim 140\degr$ in longitude.

\begin{figure}
    \centering
    \includegraphics[width=\linewidth]{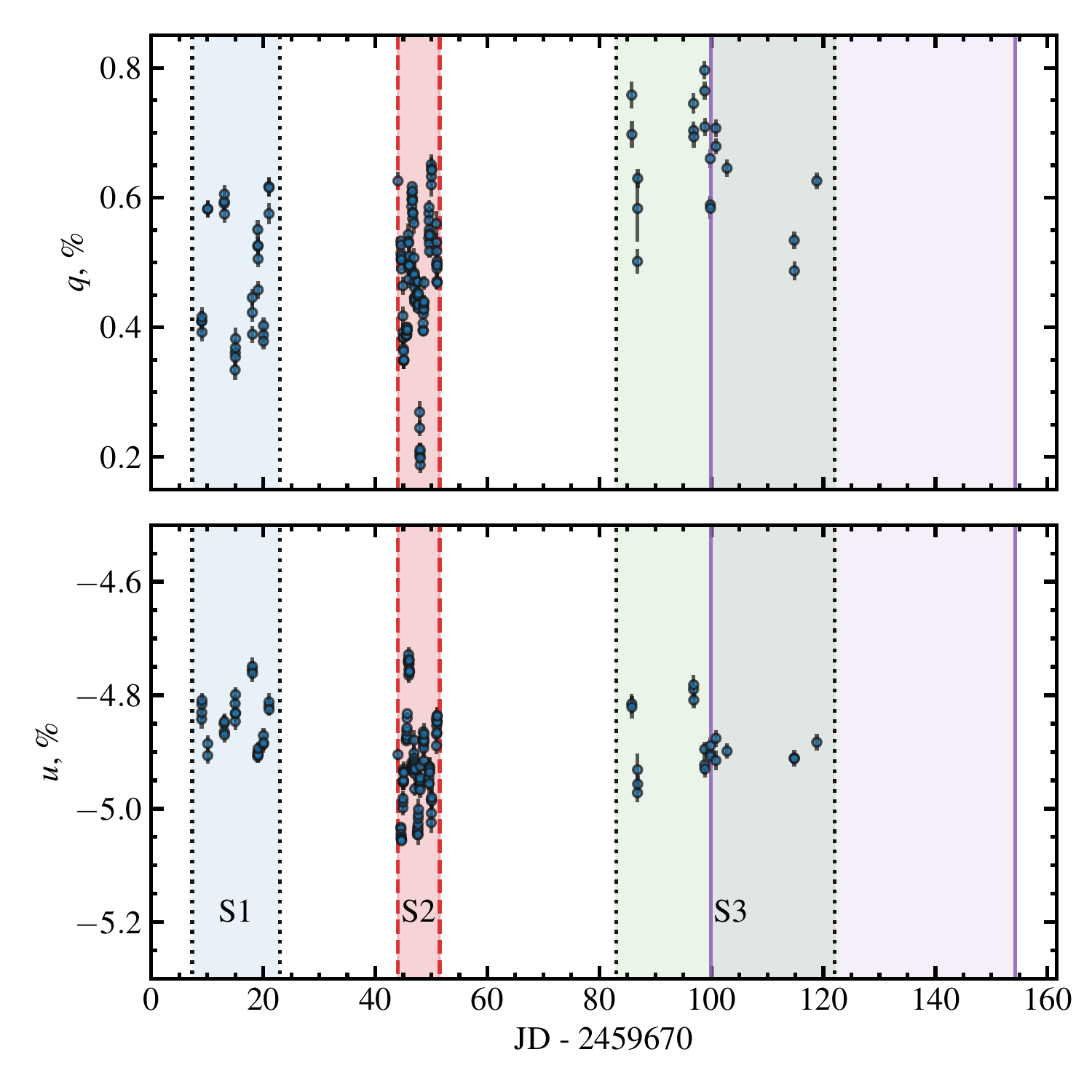}
    \caption{Observed normalized Stokes parameters for \cyg in the \textit{B} band. The vertical black dotted lines limit the observational seasons S1 and S3. The two vertical red dashed lines  show the start and the end of the IXPE campaign on 2022 May 15--21 (Season 2). The vertical purple solid lines show the start and the end of the TESS observations.}
    \label{fig:timeline_our}
\end{figure}

In this paper, we present the results of joint analysis of the new high-precision optical polarimetric observations of \cyg, historical polarimetric data obtained in 1975--1987, and optical flux measurements of TESS, which allowed us to make a new attempt to qualitatively and quantitatively constrain the geometry of \cyg.
The paper is organized as follows.
In Sect.~\ref{sec:data}, we describe the observations of the source as well as of the field stars to determine the contribution of the interstellar polarization.
In Sect.~\ref{sec:results}, we present the main results of the study: the variations of the polarization on different timescales. 
We present a model for the observed orbital and superorbital variability of the polarization in Sect.~\ref{sec:model}. 
Finally, we summarize our findings in Sect.~\ref{sec:sum}.

\section{Data acquisition and analysis}
\label{sec:data}

\begin{table}   
\caption{Log of polarimetric observations of \cyg.}             
\label{table:observations}      
\centering                          
\begin{tabular}{c c c c}        
\hline\hline                 
 Dates  & MJD & $N_\text{obs}$ & Telescope\\    
\hline                        

1975 Jun -- 1987 Oct & 42572 -- 47068 & 1511 & PMO\\
2002 April -- May     & 52381 -- 52394 & 10   & KVA\\
2002 April -- May     & 52391 -- 52394 & 4    & NOT\\
2022 March -- July  & 59652 -- 59788 & 27 & T60\\
2022 May 15--21     & 59714 -- 59719 & 6 & NOT\\
\hline                                   
\end{tabular}
\end{table}

\begin{figure}
\centering \includegraphics[width=0.9\linewidth]{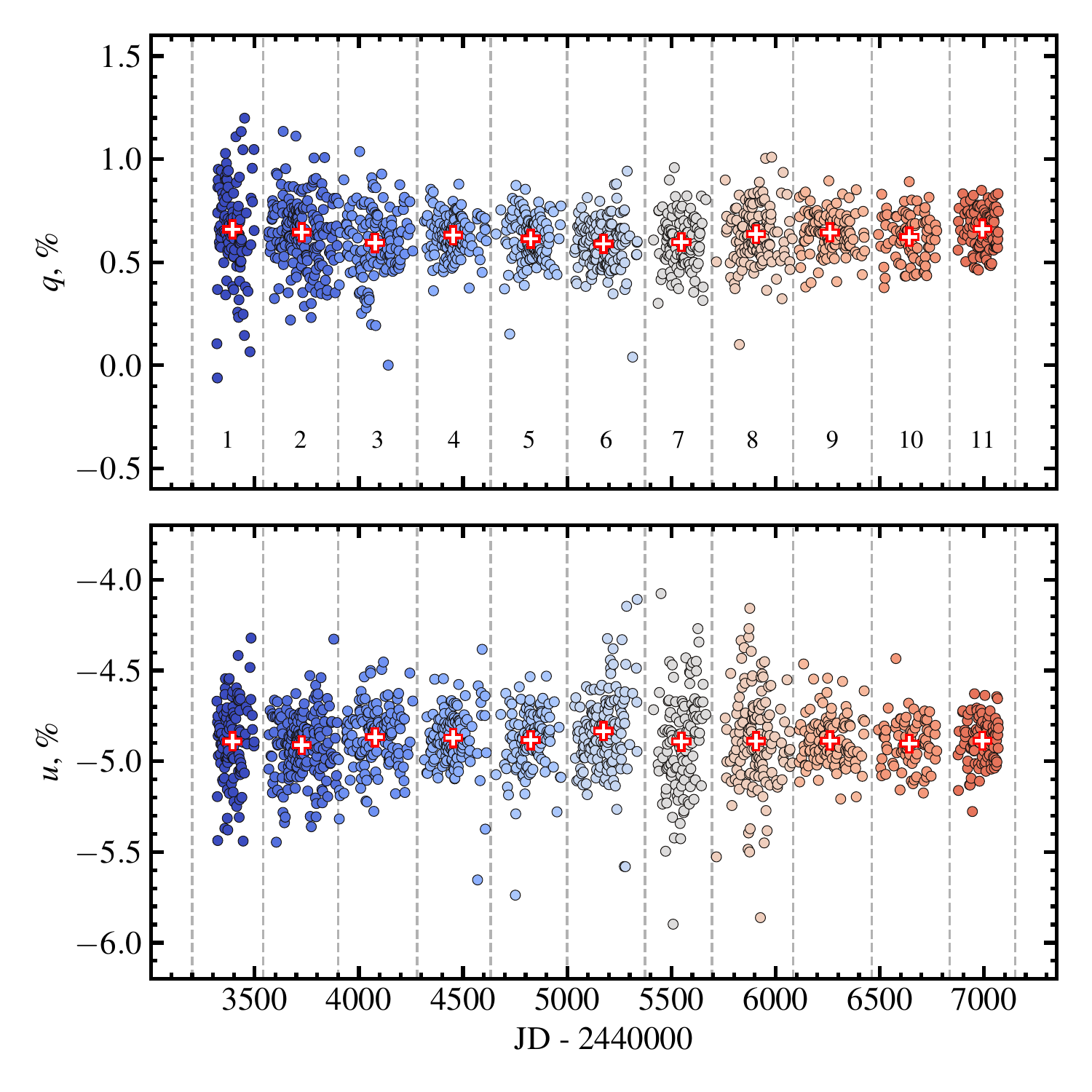}
\caption{Long-term variations of normalized Stokes parameters $q$ and $u$ of \cyg, measured in 1975--1987. The vertical dashed lines separate 11 observing seasons, roughly equal to one year of observations. Red crosses show season-averaged values. }
\label{fig:kemp_all_data_v}
\end{figure}

\begin{figure}
    \centering
    \includegraphics[width=\linewidth]{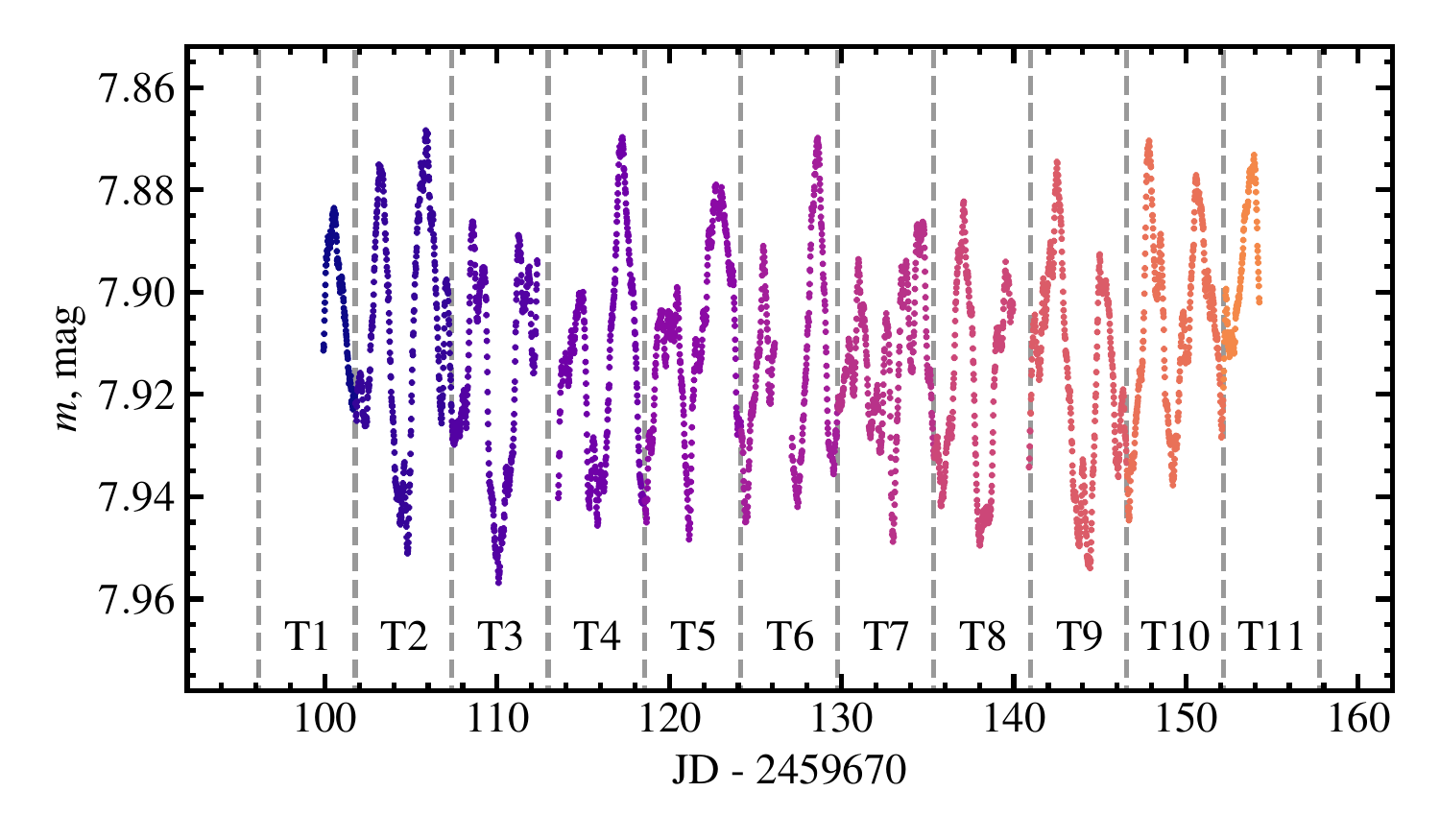}
    \caption{TESS optical light curve of \cyg. The vertical dashed lines separate consecutive orbital periods T1--T11.}
    \label{fig:tess_timeline}
\end{figure}

\begin{figure}
    \centering
    \includegraphics[width=0.8\linewidth]{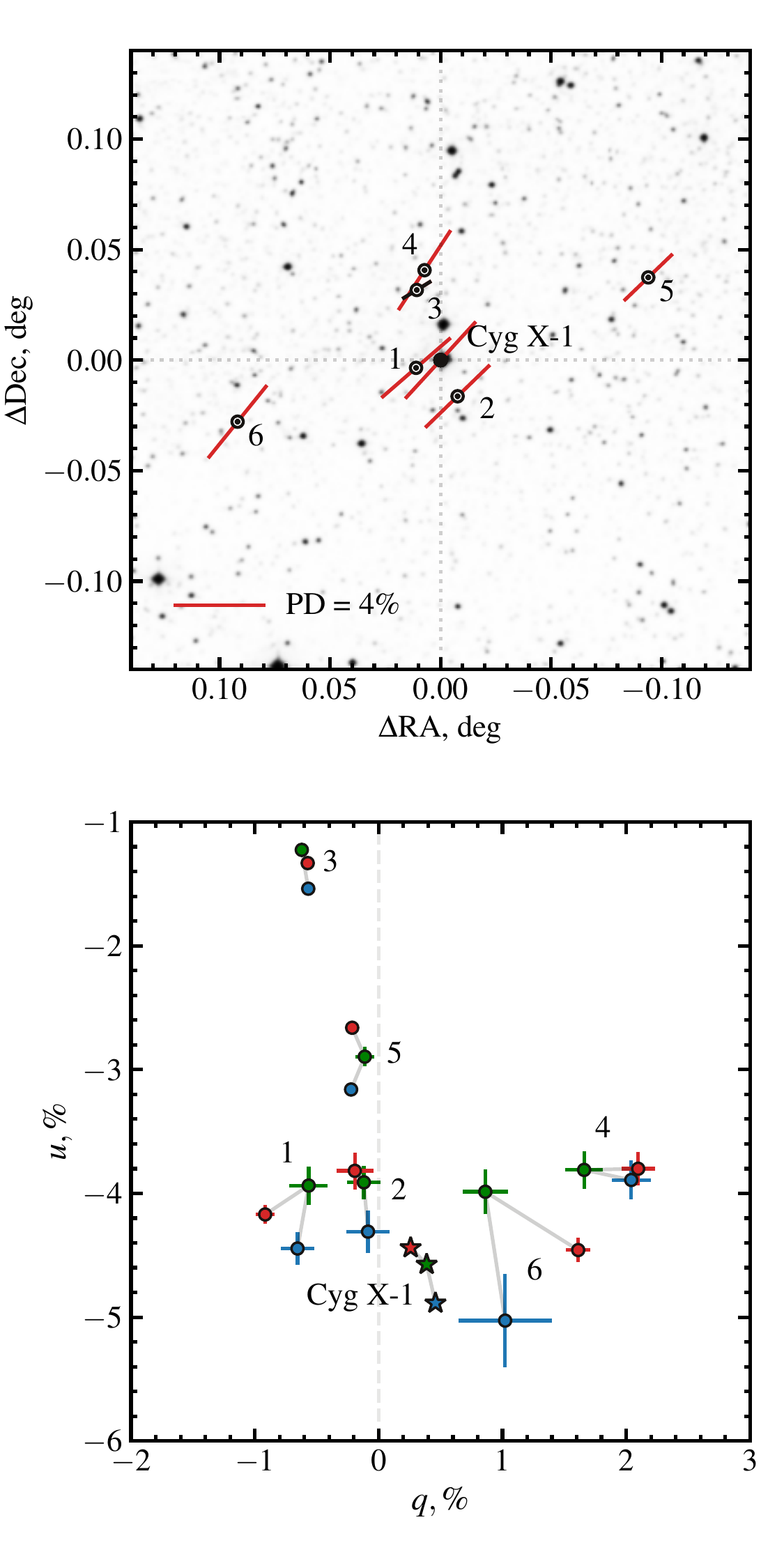}
    \caption{\textit{Top panel}: Polarization map for \cyg and field stars. The length of the bars corresponds to the PD, and the direction corresponds to the PA (measured from north to east). \textit{Bottom panel}: The observed normalized Stokes parameters $q$ and $u$ for \cyg (stars) and field stars (circles). The blue, green, and red points with 1$\sigma$ error bars correspond to \textit{B}, \textit{V}, and \textit{R} filters, respectively.}
    \label{fig:field_stars}
\end{figure}

\begin{figure}
\centering
\includegraphics[width=0.9\linewidth]{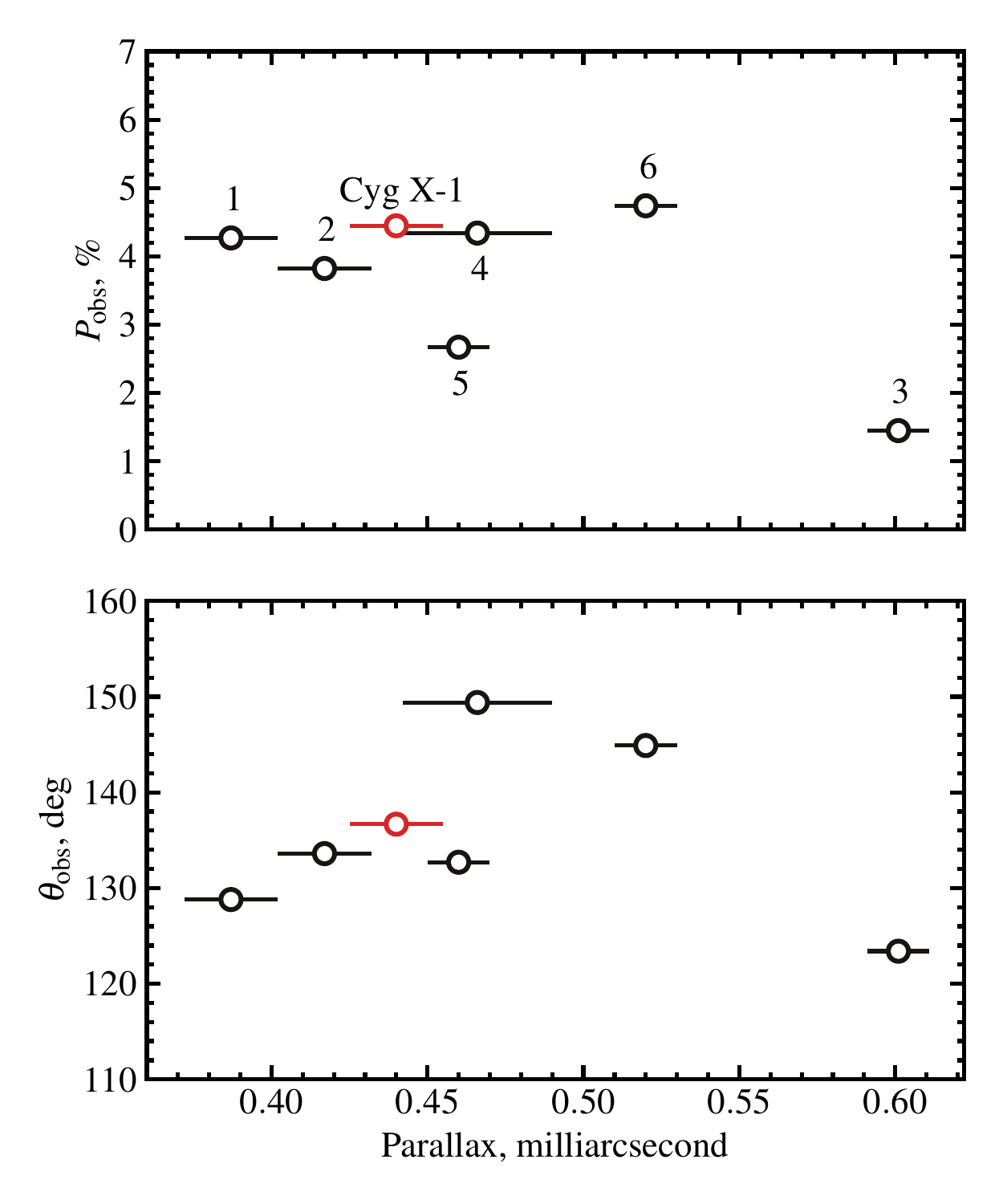}
\caption{Dependency of PD (\textit{top panel}) and PA (\textit{bottom panel}) on Gaia parallax for \cyg (red circle) and field stars (black circles) in $R$ band. The horizontal error bars correspond to the errors on Gaia parallaxes. The errors in the PD and PA are smaller than the symbol size. }
\label{fig:polar_parallax}
\end{figure}

We performed high-precision optical polarimetric observations of \cyg with the broad-band \textit{BVR} polarimeters \DP \citep{Piirola2014} and \DUF \citep{Piirola2021}.
\DP is mounted on the remotely-controlled 60~cm Tohoku telescope (T60) at Haleakala Observatory, Hawaii.
\DUF is a visitor instrument installed at the 2.56~m Nordic Optical Telescope (NOT), Observatorio del Roque de los Muchachos (ORM), La Palma, Spain.
\DP and \DUF are high-precision double-image CCD polarimeters, capable of measuring polarization simultaneously in three optical (\textit{BVR}) bands. 
The polarization of the sky (even if it changes during observations) is optically eliminated by the design of the instruments.
The instrumental polarization of both instruments is small ($<10^{-4}$) and is well calibrated by observing 15 to 20 unpolarized standard stars.
The zero point of the PA was determined by observing highly polarized standards HD~204827 and HD~161056.
Each measurement of Stokes parameters (\qobs, \uobs) took about 20~s and more than 200 individual measurements were obtained during the average observing night.
A more detailed description of the methods and calibrations can be found in \citet{Piirola2020} and \citet{Kravtsov2022}.

Cyg X-1 was observed for 27 nights between 2022 March 28 and July 28 at the T60 and for 6 nights, 2022 May 15--21 at the NOT (see Table~\ref{table:observations}).
Taking advantage of the $\sim$140\degr\ difference in the longitude of the observatories, we covered 30\% of the orbit of \cyg simultaneously with IXPE observations (red stripe in Fig.~\ref{fig:timeline_our}).
We refer to this data set as Season 2 (or just S2) hereafter.
The typical errors on the 30-minute-averaged measurement of Stokes parameters are $\sigma_{\rm p} \approx 0.004\%$ for the NOT and $\approx 0.01\%$ for the T60 data.

\begin{table*}   
\caption{Observed PD and PA, interstellar polarization, and intrinsic polarization of \cyg, obtained by averaging S1--S3 data.  
}             
\label{table:polar}      
\centering                          
\begin{tabular}{ccccccccc}        
\hline\hline                 
  & \multicolumn{2}{c}{Observed} & & \multicolumn{2}{c}{Interstellar} & &  \multicolumn{2}{c}{Intrinsic}  \\  
\cline{2-3} 
\cline{5-6}
\cline{8-9}
Filter & $P_{\text{obs}}$ & $\theta_{\text{obs}}$ & & $P_{\text{is}}$  & $\theta_{\text{is}}$ && $P_{\text{int}}$ & $\theta_{\text{int}}$  \\    
  &  (\%) &  (deg) &&  (\%) &   (deg) && (\%) &   (deg) \\  
\hline                        
   \textit{B}  & $4.91 \pm 0.06$ & $138 \pm 1$ && $4.31 \pm 0.17$ & $134 \pm 1$ && $0.83 \pm 0.17$ & $158 \pm 6$ \\ 
   \textit{V}  & $4.59 \pm 0.06$ & $138 \pm 1$ && $3.91 \pm 0.14$ & $134 \pm 1$ && $0.86 \pm 0.14$ & $155 \pm 5$ \\
   \textit{R}  & $4.44 \pm 0.05$ & $137 \pm 1$ && $3.82 \pm 0.15$ & $134 \pm 1$ && $0.79 \pm 0.15$ & $155 \pm 5$ \\ 
\hline                                   
\end{tabular}
\end{table*}

We additionally use the historical observations carried out at the Pine Mountain Observatory (PMO), United States, in 1975--1987 (see Table~\ref{table:observations} and Fig.~\ref{fig:kemp_all_data_v}). 
The reduced data were presented in parts in several papers \citep{Kemp1978, Kemp1979, Kemp1983}, but have never been published as a raw dataset.
With the kind permission of the Pine Mountain Observatory staff, we are publishing these data (hereafter PMO data) in the public domain for the first time.\footnote{All the polarimetric data we used in the paper is available in VizieR.}
We refer to these observations as PMO data hereafter.
The description of the observational techniques can be found in \citet{Kemp1981}. Each PMO observation is a nightly average value with a typical integration time of several hours.  
We also use previously unpublished polarimetric observations of \cyg, carried out in April--May 2002 with the TurPol polarimeter \citep{Piirola1973, Piirola1988}, installed on NOT and 60~cm KVA telescope, ORM, La Palma.

In addition to the polarimetric data, we retrieved and analyzed the publicly available\footnote{\url{https://archive.stsci.edu/missions-and-data/tess}} Transiting Exoplanet Survey Satellite \citep[TESS;][]{TESS} optical light curves of \cyg.
We used 2-minute cadence calibrated data (\texttt{PDCSAP\_FLUX}), obtained in sectors 54 and 55 (July -- September 2022, see Fig.~\ref{fig:tess_timeline}).

\begin{figure}
\centering
\includegraphics[width=\linewidth]{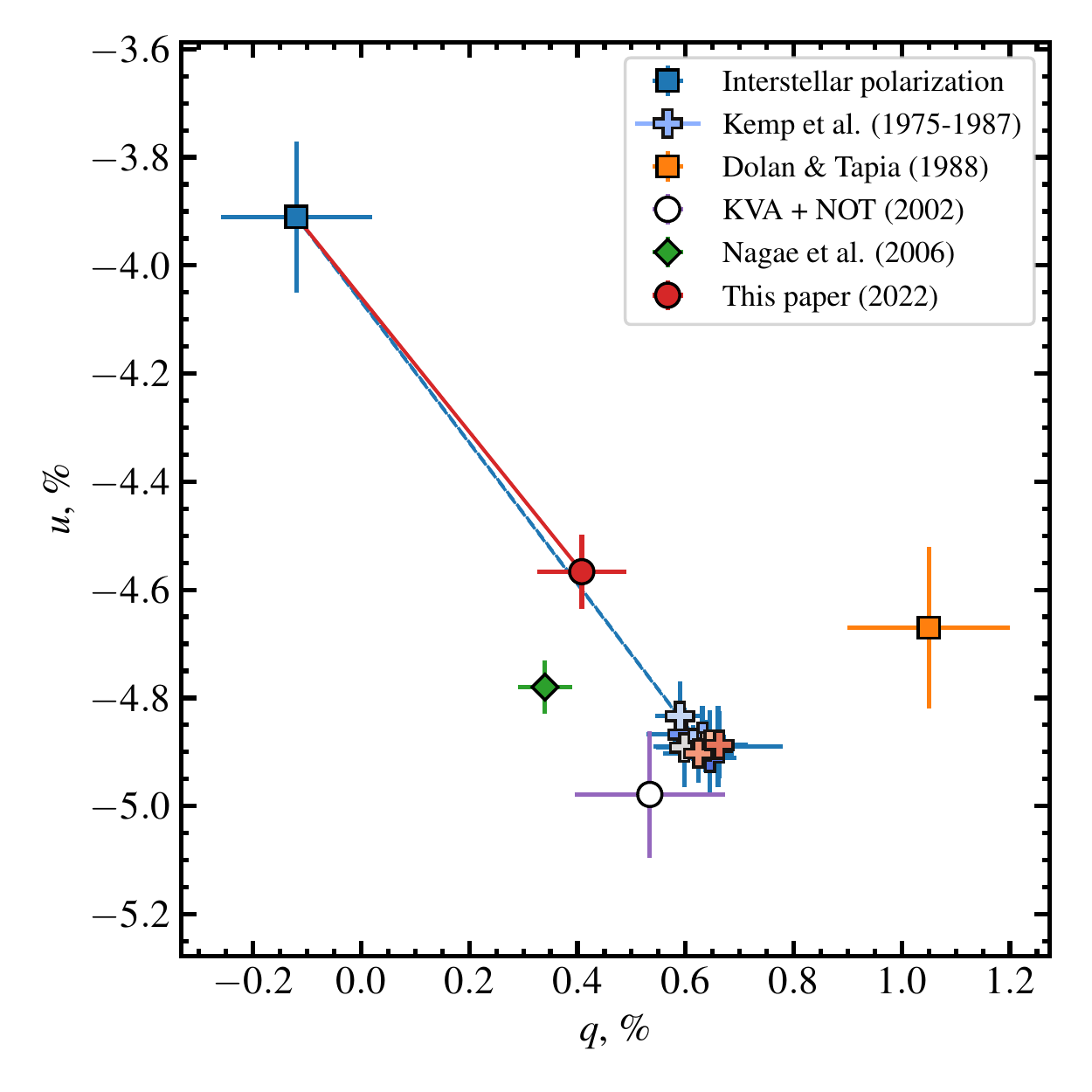}
\caption{Normalized Stokes parameters (\qobs, \uobs) of \cyg in the \textit{V} band. The red circle with the error bar is the average polarization of \cyg in 2022. The crosses of different colors (from cool to warm) with error bars correspond to the average polarization in each of the 11 seasons of PMO observations in 1975--1987. Blue dashed and red solid lines show the directions of the average intrinsic polarization vector of \cyg for PMO and our observations, respectively. Other symbols correspond to the data obtained in other epochs, as described in the inset.}
\label{fig:qu_plane}
\end{figure}

\section{Results}
\label{sec:results}

\subsection{Average intrinsic polarization and its secular changes}

The observed polarization of \cyg is the sum of the intrinsic polarization of the source and the interstellar (IS) polarization component, that arises from the dichroism of the dust grains located between the observer and the target.
The IS polarization is estimated and subtracted from the observed polarization.
To find a reliable estimate of IS polarization in the source direction, we observed a sample of 6 field stars (see Fig.~\ref{fig:field_stars}), which are close in distance to Cyg X-1 as indicated by their Gaia parallaxes (see Fig.~\ref{fig:polar_parallax}).
We consider the wavelength dependence of the observed polarization (to exclude stars with intrinsic polarization) and take into account both the angular separation and proximity to the target. 
We conclude that the polarization of the star Ref~2 from our sample can serve as the IS polarization estimate.
Hereafter we denote the normalized Stokes parameters of Ref~2 as (\qis, \uis) and subtract them from the observed values of the target (\qobs, \uobs) to obtain Stokes parameters of the intrinsic polarization (\qint, \uint).
These are translated to the intrinsic PD $P$ and angle $\theta$
\begin{equation}
P = \sqrt{q_\text{int}^2 + u_\text{int}^2},\qquad 
\theta = \frac{1}{2}\arctantwo(u_\text{int}, q_\text{int}).  
\end{equation}
The uncertainty on the PD is equal to the uncertainty of the individual Stokes parameters, while uncertainty on the PA in radians was estimated as $\sigma_{\theta} = \sigma_{\rm p}/(2P)$ \citep{Serkowski1962,Kosenkov2017}. 
The average observed and intrinsic polarization of \cyg, as well as the interstellar polarization estimates, are listed in Table~\ref{table:polar}.

A big fraction of the observed polarization, about 4\% out of a total 4.5--5\%, has IS origin \citep{Kemp1979, Nagae2009}. 
Subtracting the interstellar component from the observed polarization, we find the intrinsic PD of $0.8\%\pm0.2\%$ with the PA of $155\degr\pm5\degr$  (or equivalently $-25\degr$; see Table~\ref{table:polar}).
This value is comparable to the characteristic optical PDs in other accreting BH X-ray binaries in the outburst \citep{Kosenkov2017,Veledina2019}. 
The average intrinsic optical PA   matches within the errors that measured in the X-rays \citep[$\theta_\text{X} = -22\fdg2\pm2\fdg2$,][]{Krawczynski2022}.

The uniquely long history of polarimetric studies of Cyg X-1 allows us to track the long-term evolution of the average PD almost 50 years back.
We split the PMO $V$-band observations into eleven bins (PMO1--PMO11, see Fig.~\ref{fig:kemp_all_data_v}), each about a year long and calculate the average values of the observed Stokes parameters within each bin.
We plot them in the ($q$, $u$)-diagram in Fig.~\ref{fig:qu_plane} (colored crosses) along with our NOT+T60 2022 average polarization (red circle), KVA+NOT 2002 data (empty circle), and other published data \citep[orange square and green diamond,][]{Dolan1989, Nagae2009}.
We show the estimated value of IS polarization with a blue square.

Blue and red lines connect the PMO and NOT+T60 2022 data with the IS estimate, respectively.
The length and direction (from IS estimate towards the data points) of these lines correspond to the vectors of the average intrinsic polarization for different epochs.
The vector directions match with high accuracy ($\Delta \theta_\text{int} < 1\degr$).
This supports our choice of the reference star Ref 2 as an estimate of interstellar polarization, as the alignment of the intrinsic polarization vectors is unlikely to be accidental.
We note that the other historical values shown in Fig.~\ref{fig:qu_plane} are substantially scattered in the $q-u$ plane, despite their small error bars. 
This may be caused by the orbital variations -- at least several different orbital periods must be averaged to get a robust estimate of the average polarization.

The average intrinsic polarization for our NOT+T60 2022 data differs significantly from the PMO data ($|\Delta P_{\rm int}| \approx 0.4\%$), indicating the presence of secular changes in the PD meanwhile preserving a constant PA.
The decrease of the intrinsic PD may be caused by the decrease of the scattered flux, tailored to the secular changes of the wind density, changes of the accretion disk size, or/and its spatial orientation.
In Fig.~\ref{fig:kemp_all_data_v} we see that the one-year polarization averages change slightly from season to season. 
Different spread of all data points shows that the amplitude of the variability also varies on yearly timescales. 

\subsection{Short-term variability of orbital profiles}

\begin{figure}
    \centering
    \includegraphics[width=\linewidth]{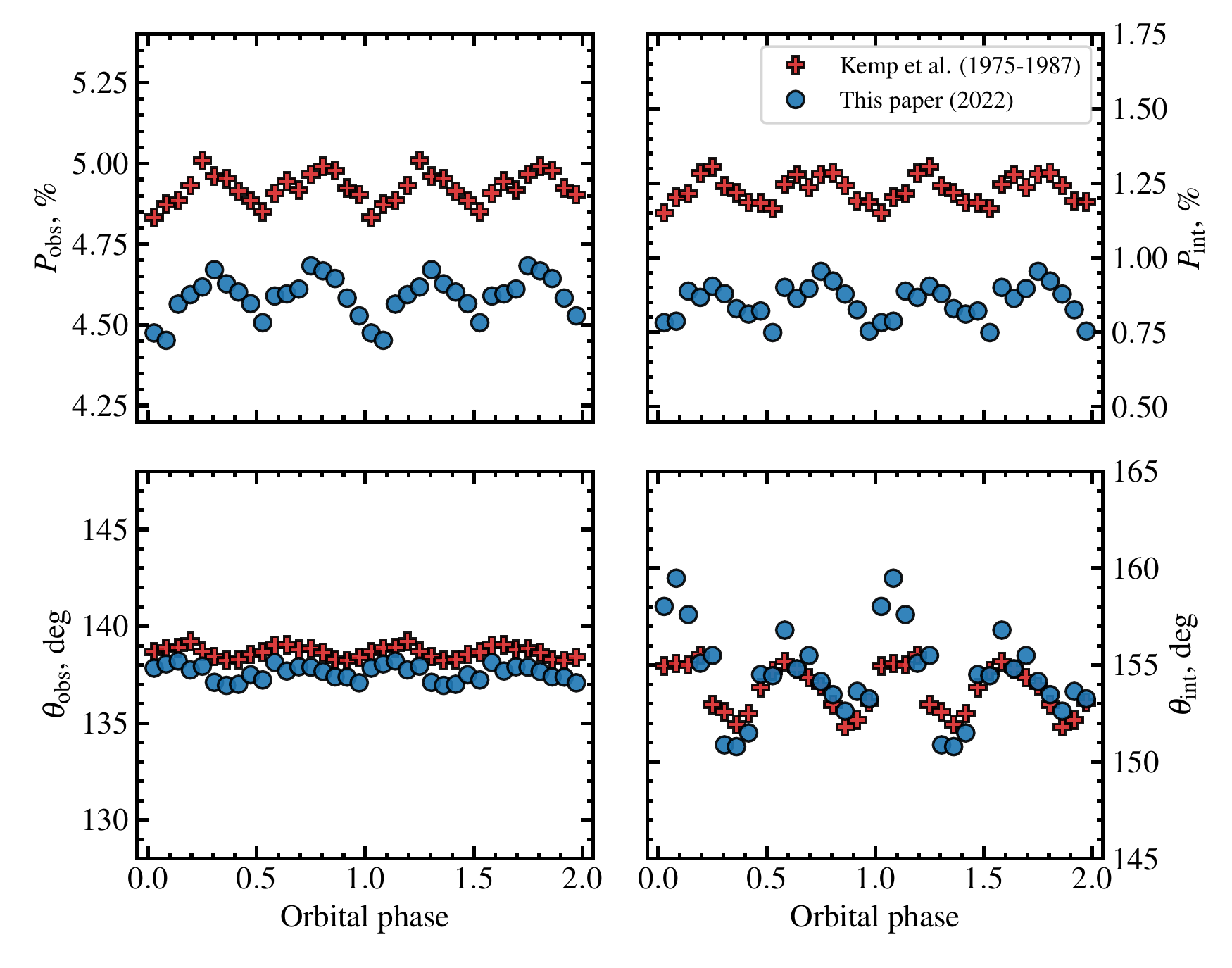}
    \caption{\textit{Left panels}: Observed PD, $P$, (\textit{top panel}) and PA, $\theta$, (\textit{bottom panel}) of \cyg in the \textit{V} band, folded with the orbital period. The blue circles correspond to the data, obtained in 2022. The red crosses correspond to the PMO data, obtained in 1975--1987. Each point corresponds to the average value, calculated within the phase bin of width $\Delta \phi = 1/18$. The typical $1\sigma$ uncertainty is smaller than the symbol size.
    \textit{Right panels:} same as left, but intrinsic polarization of \cyg is shown.}
    \label{fig:Kemp_and_our_p_pa}
\end{figure}

\begin{figure}
    \centering
    \includegraphics[width=0.93\linewidth]{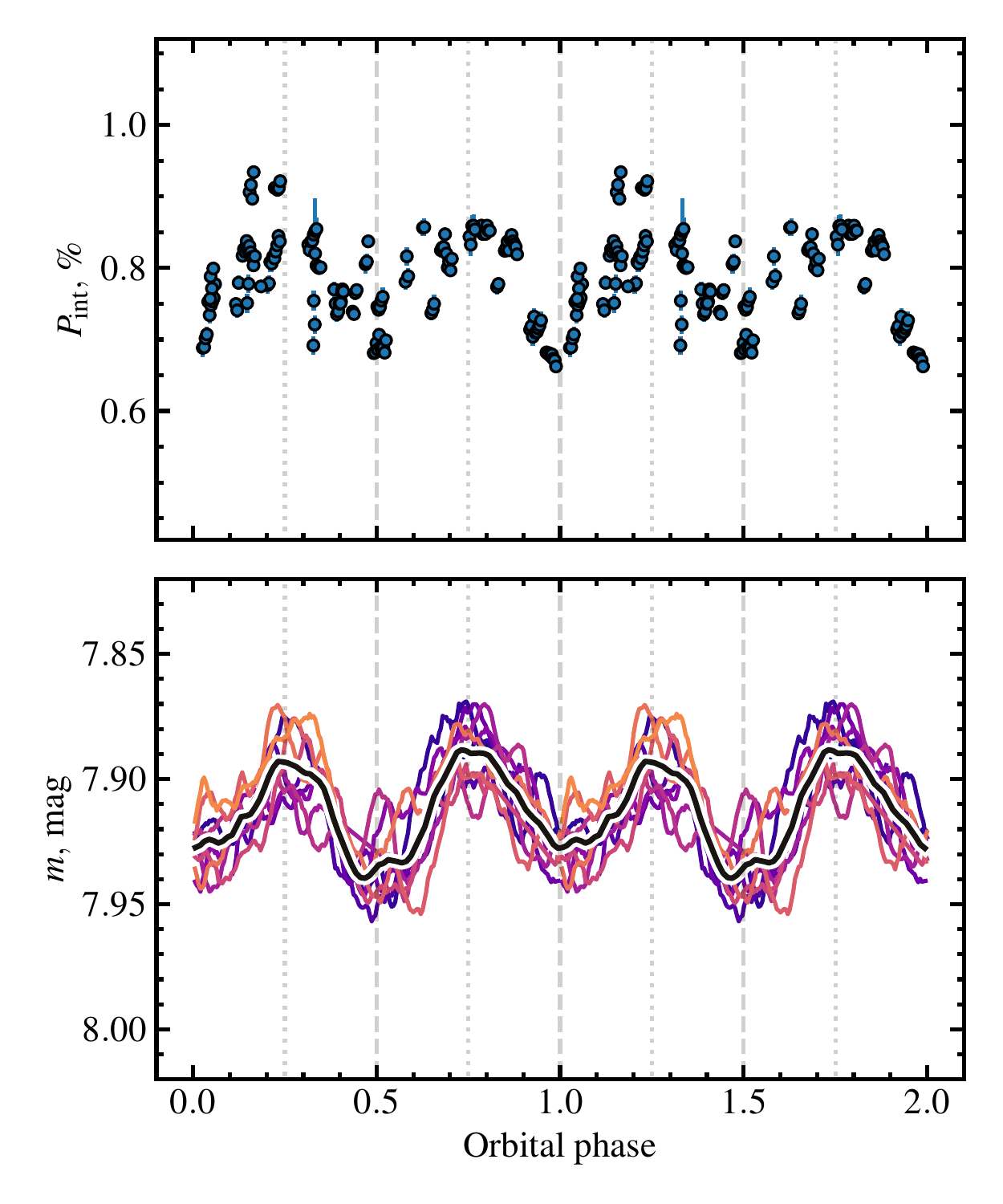}
    \caption{\textit{Top panel:} Intrinsic polarization of \cyg in the \textit{R} filter, folded with the orbital period (Seasons 1--3 are plotted). Each circle with $1\sigma$ error bar shows the average polarization, calculated within the 30-minute bin.  \textit{Bottom panel:} TESS magnitude of \cyg, folded with the orbital period. Different colors (from cold to warm) correspond to different orbital periods T1--T11. The black solid line shows the average orbital profile. }
    \label{fig:tess_and_pol}
\end{figure}

The significant variability of Stokes (\qobs, \uobs) parameters with the amplitude of about 0.1\%--0.2\% can clearly be seen in our 2022 observations (see Fig.~\ref{fig:timeline_our}).
We performed a timing analysis of our $BVR$ polarimetric data which revealed that the main period of variations of the Stokes parameters has not changed since the 1970s and is equal to half of the orbital period within errors.
To study the possible changes in the average orbital profiles of the polarization over decades, we folded our and PMO polarization data with the orbital phase, adopting the period $P_{\rm orb}$ from the photometric ephemeris \citep{Brocksopp1999a}.
To suppress the stochastic and instrumental noise, we split the data into 18 orbital phase bins, in which we calculate weighted average values of Stokes parameters that are subsequently used to obtain the PD $P$ and PA $\theta$.
The comparison between our and PMO polarization light curves is shown in Fig.~\ref{fig:Kemp_and_our_p_pa}. 
Except for the systematic offset between our and PMO data the nature of which is discussed above, the shapes of the average PD and PA orbital profiles are in exceptionally good agreement with each other.

To check how the shape of optical light curves changes from one period to another, we split TESS photometric data into eleven consecutive orbital periods T1--T11 (see Fig.~\ref{fig:tess_timeline}).
The shape of each individual profile is far from a double-sine wave expected in the case of ellipsoidal variations caused by the rotation of a tidally distorted star around the center of mass.
A short-period variability is superimposed on the main double-sine curve, which leads to changes in the amplitudes and phases of the main maxima/minima.
In Fig.~\ref{fig:tess_and_pol} we show the orbital profiles of the intrinsic polarization in \textit{R} band together with TESS photometric profiles.
Both optical polarization and optical flux show the double-sinusoidal orbital variations with the minima in the conjunctions (phases 0 and 0.5) and maxima in the quadratures (phases 0.25 and 0.75). 

Photometric variations arise from the non-spherical shape of the tidally distorted companion, whose visible area (and hence the flux) is maximal around the quadratures.
If the scattering of the donor star emission occurs in a region connected to the compact object, we expect the scattering angle to reach a maximum of $90\degr$ at the same phases, corresponding to the maximum PD in the case of Thomson scattering.
In the conjunctions, the visible area of the supergiant approaches its minima resulting in the minima in flux, while at the same time the scattering angle reaches minimum or maximum leading to a smaller PD.
The short-term changes in both the flux and polarization, which are superimposed on the periodic variations, can originate from one or more mechanisms -- the pulsations of the main star, the presence of the spots on its surface, the inhomogeneities of the wind, or the eclipses of the bright parts of the disk by the infalling matter. 
We note that despite the correlation of flux and polarization, the polarization variability cannot be explained only by variations of the unpolarized flux. Indeed, for the intrinsic PD $P = F_{\rm pol}/F_{\rm tot} \sim 0.01$ the unpolarized flux variations of the order of  $\Delta F_{\rm tot} \sim 5 \times 10^{-3}F_{\rm tot}$ gives negligibly small polarization variability:
$$\Delta P  \approx P \frac{\Delta F_{\rm tot}}{F_{\rm tot}} \sim 5 \times 10^{-5},$$
while the observed one is at least factor 20 of larger.

\begin{figure}
    \centering
    \includegraphics[width=0.85\linewidth]{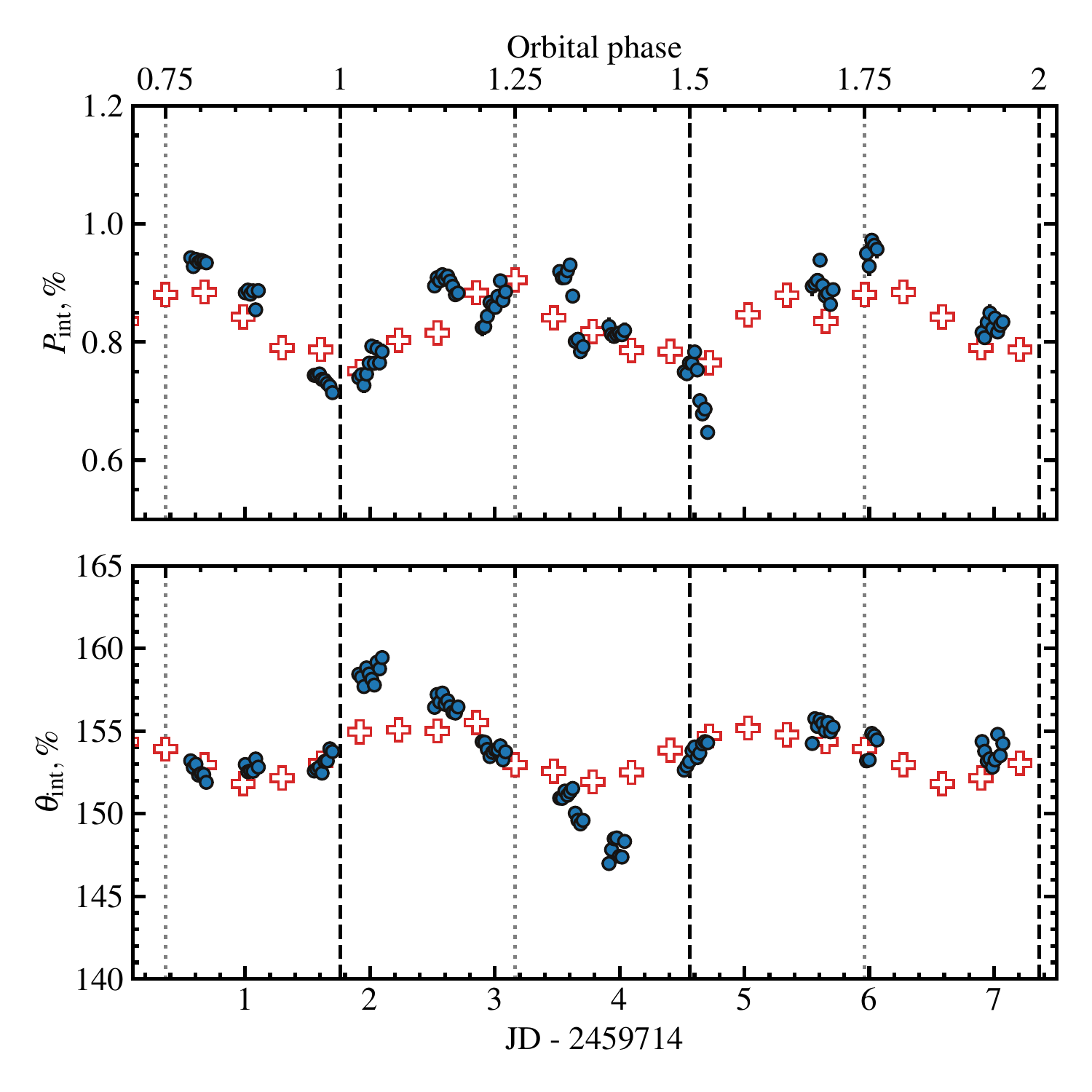}
\caption{Variability of the intrinsic PD (\textit{upper panel}) and PA (\textit{bottom panel}) of \cyg in \textit{V} band, measured in May 2022 (blue circles). Each circle corresponds to the average value, calculated within the 30-minute bin. The $1\sigma$ errors are smaller than the symbol size. The red crosses correspond to the average binned polarization, measured by Kemp with colleagues during 1975--1987 with the constant shift in PD by $\Delta P = -0.4\%$.}
    \label{fig:intrinsic_set2}
\end{figure}

With our exceptionally dense orbital coverage, we can compare the profiles of a single cycle with the average one, as given by the PMO data.
In Fig.~\ref{fig:intrinsic_set2} we show the profile obtained from the 8-days-long monitoring of \cyg during Season~2, overlaid on top of the average profile of the polarization variability in the $V$ band.
As we see from the figure, although the overall shapes of the polarization variability curves are roughly consistent with 1975--1987 patterns, the amplitude of variations is substantially higher in our Season~2 data, where the harmonic content is also richer.
These facts support the statement of \citet{Dolan1989} on the existence of non-orbital polarization variability and the importance to account for it when extracting orbital parameters.
In the following sections, we describe the modeling of the polarization variability curves with different analytical models and discuss how the short-term variability affects the results. 

\subsection{Superorbital evolution of polarization profiles}

In addition to the short-term variability, hints of the long-term changes of \cyg polarization profiles have been reported by several authors.
\citet{Kemp1983} suggested the presence of long-term optical polarization variations at the superorbital period of 294~d, discovered in the X-rays \citep{Priedhorsky1983}. 
The authors discussed several models that could explain the variations, including the precession of the accretion disk and the obscuration of the scattering medium.
Comparing the average optical polarization obtained between 1975 and 2006, \citet{Nagae2009} found secular variations of the average polarization component of \cyg.

\begin{figure}
\centering
\includegraphics[width=0.93\linewidth]{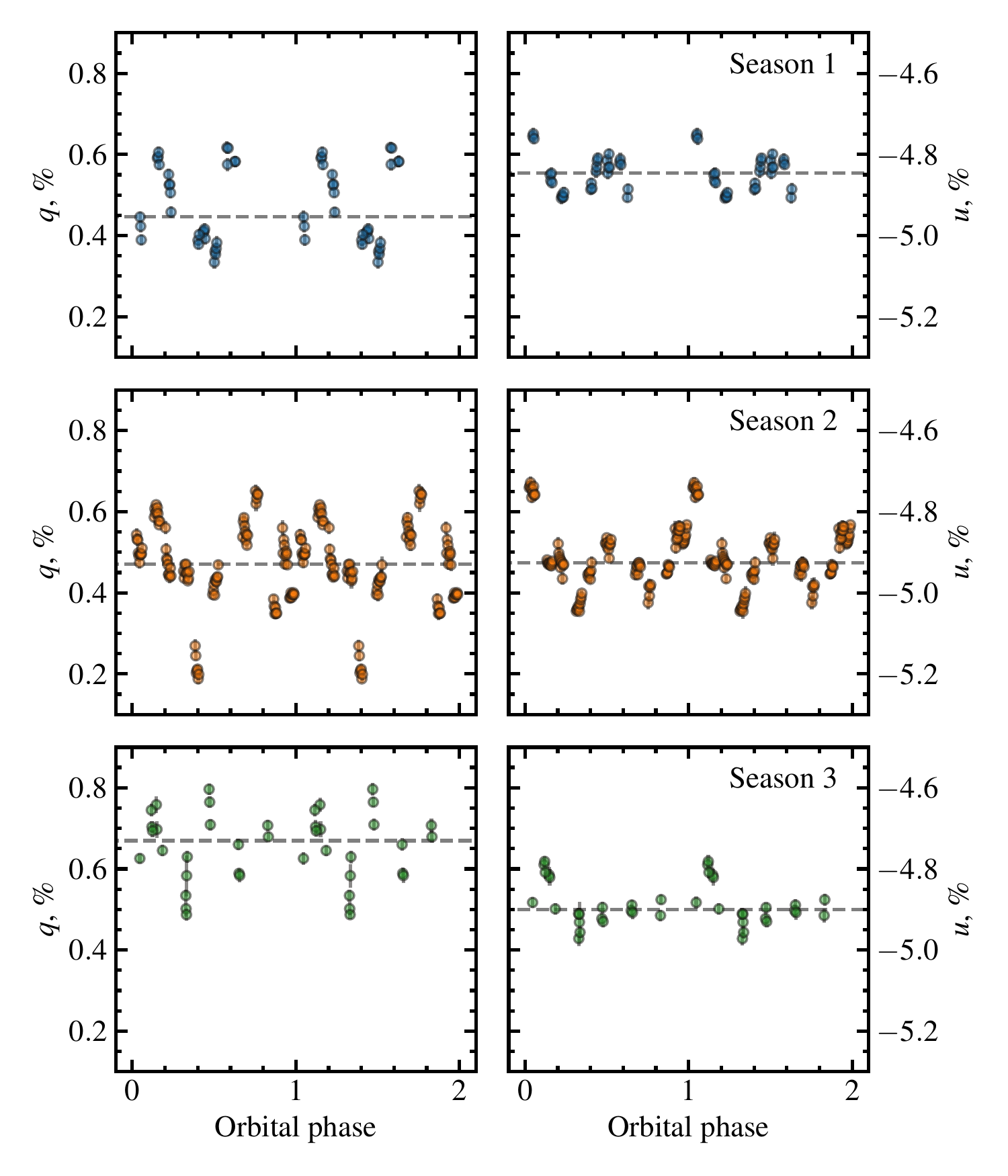}
\caption{Orbital variations of the observed normalized Stokes parameters $q$ (left) and $u$ (right) of \cyg in the \textit{B} band for different seasons (seasons 1, 2, and 3 from top to bottom). Horizontal dashed lines show the weighted average values of the corresponding parameters.}
\label{fig:Stokes_all}
\end{figure}

\begin{figure}
\includegraphics[width=0.9\linewidth]{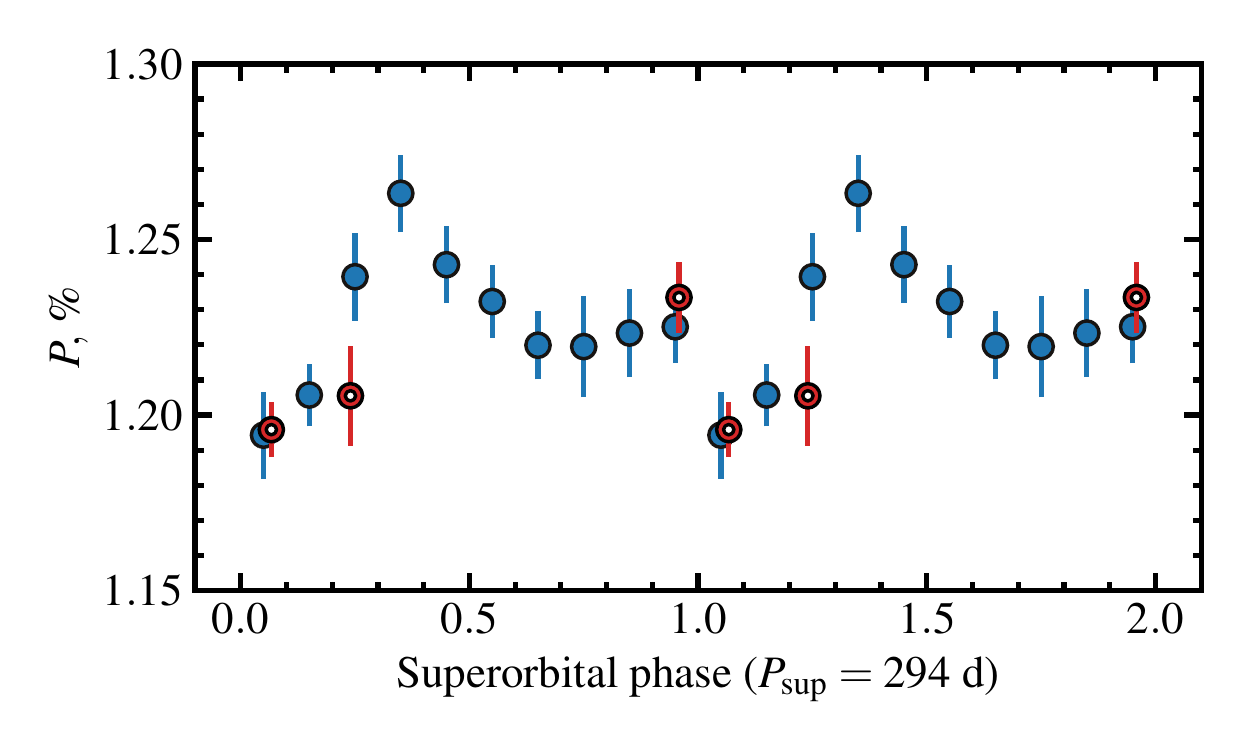}
\caption{Intrinsic PD of \cyg in the \textit{V} band for PMO (blue circles) and S1--S3 (red hollow circles) data, folded with the superorbital period. Red points are shifted by a constant $\Delta P = -0.4\%$ in the vertical direction to take into account secular changes in the PD.  }
\label{fig:superorbital}
\end{figure}

We find signatures of the long-term variability in our 2022 polarimetric data.
Fig.~\ref{fig:Stokes_all} shows how the polarization profiles change with the superorbital phase, separated roughly by a month (see Fig.~\ref{fig:timeline_our}) -- we see significant changes in the average values of Stokes parameters along with the changes in the amplitude and profiles of orbital variations. In Fig.~\ref{fig:superorbital} we show the superorbital profile of \textit{V}-band polarization of \cyg. The average values of the PD for seasons S1 -- S3 (red hollow circles), folded with the superorbital period ($P_\mathrm{sup} = 294$~d, JD$_0$ = 2440000) are consistent with the same part of the superorbital profile, observed in PMO data.

\section{Modeling}
\label{sec:model}

\begin{figure}
    \centering
    \includegraphics[width=\linewidth]{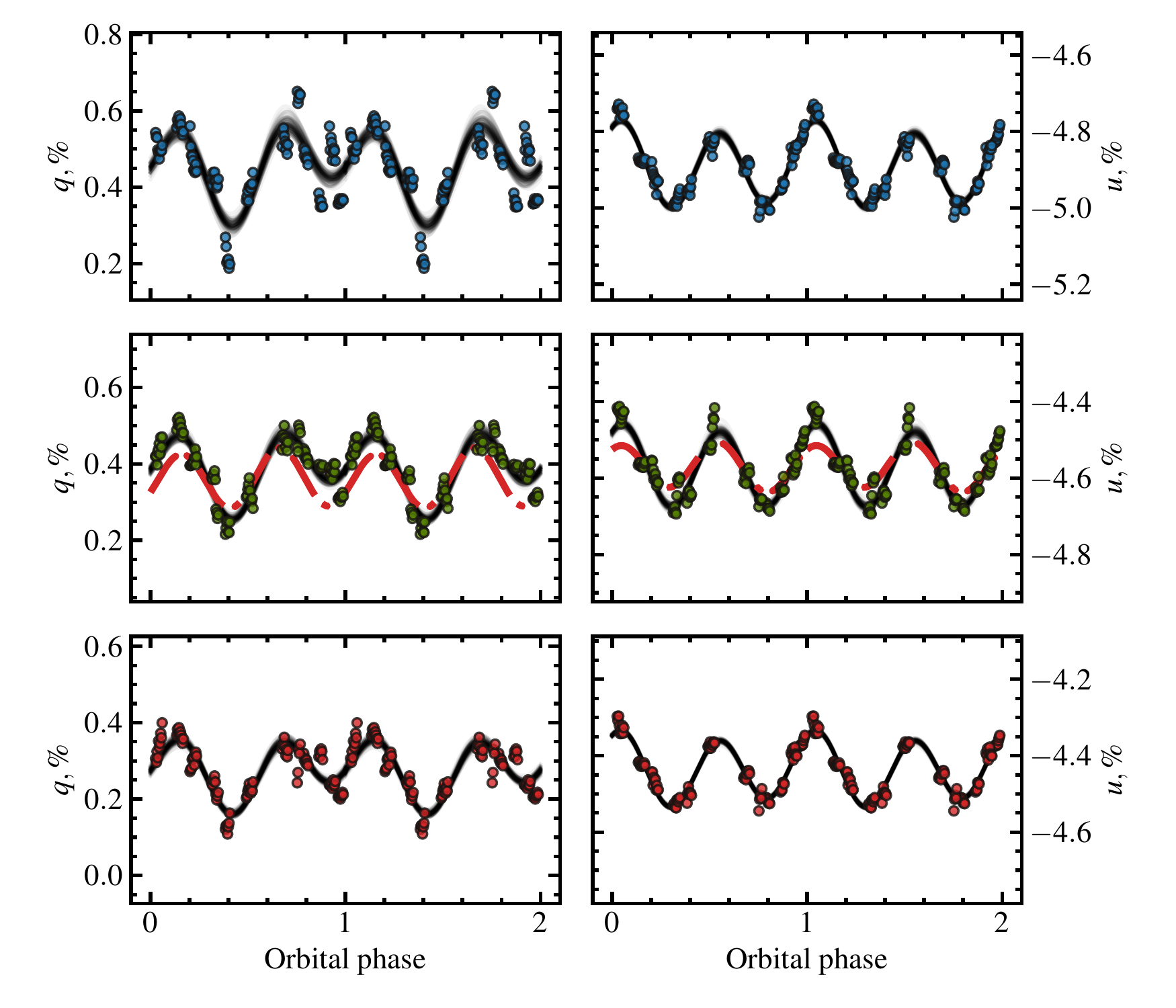}
    \caption{Variability of the observed Stokes parameters of \cyg, obtained during Season 2 in \textit{B}, \textit{V}, and \textit{R} bands (\textit{top}, \textit{middle}, and \textit{bottom panels}, respectively). Each circle corresponds to the average value, calculated within the 30-minute bin. The $1\sigma$ errors are smaller than the symbol size. The solid black lines correspond to the best fit with the Fourier series given by Eq.~(\ref{eq:Fourier}).  The dashed red lines on the middle panels show the best fit of PMO historical \textit{V} band data with the same model, shifted vertically to overlap our data. }
    \label{fig:set2_fourier_modelling}
\end{figure}

To explain the behavior of polarization at different time scales, we consider several possibilities for the geometry of scattering matter.
We start with the generic model for polarization production in binary systems, where it arises from the Thomson scattering of the companion star radiation by a cloud of an optically thin matter near the compact object \citep{BME}.
The key assumption of the model is the co-rotation of the scattering material with the secondary -- in our case, with the compact object.
The PD in this case peaks at orbital phases where the scattering angle is $90\degr$. 
For a circular orbit, Stokes parameters of the linear polarization vary as a sine-like wave at a double orbital frequency.
In the case of eccentric orbit or/and asymmetry of the distribution of the light-scattering material about the orbital plane, the profiles become skewed and can be described by adding the first harmonic of the orbital period. 
Alternatively, the appearance of the first harmonic can be related to the presence of an optically-thick scattering material. 
Below we study the harmonic content of polarization profiles and consider different possibilities for the geometry of the scattering matter.

\subsection{Fourier method}

\begin{figure}
    \centering
    \includegraphics[width=0.9\linewidth]{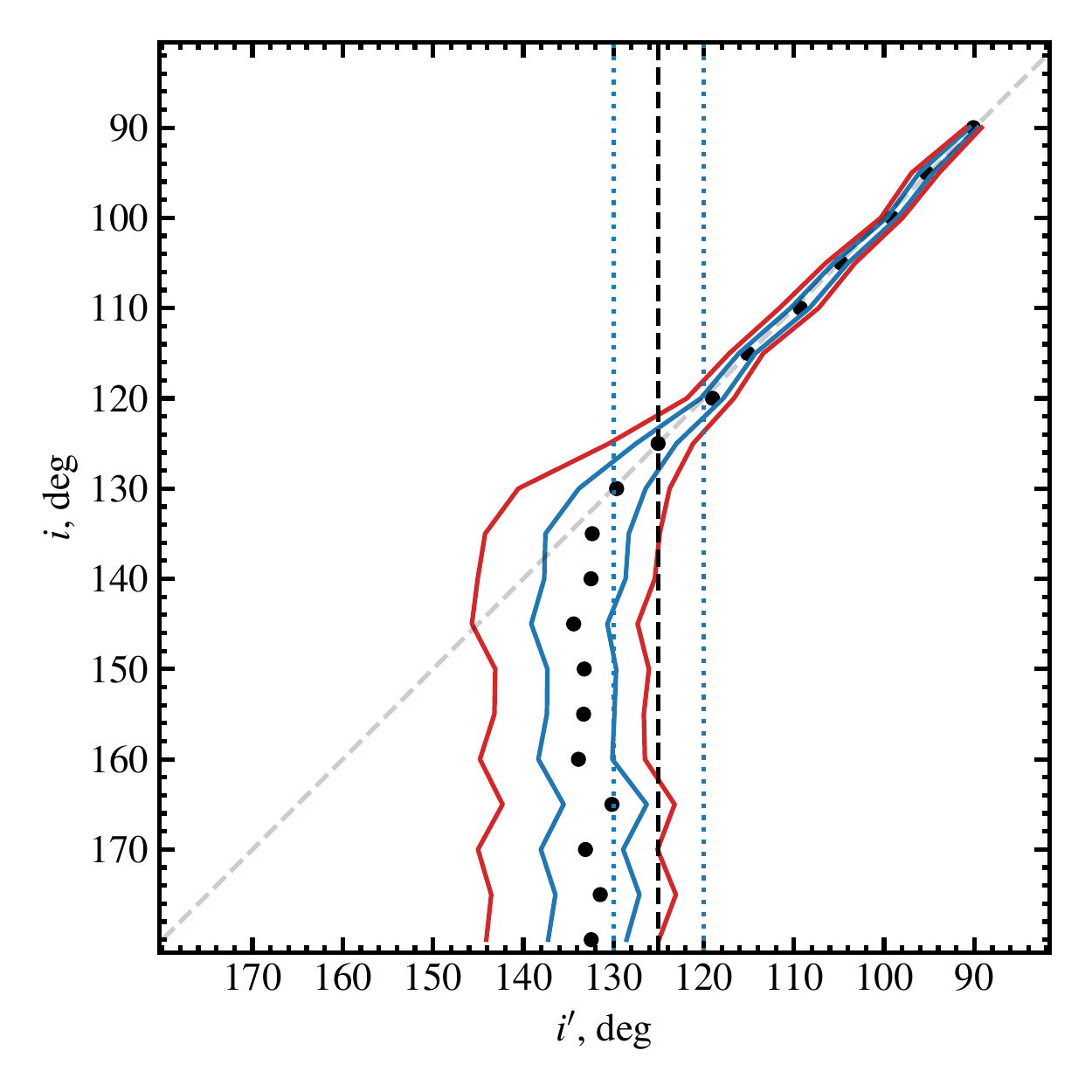}
    \caption{Estimated $1\sigma$ and $2\sigma$ confidence intervals on the true inclination $i$ for given estimate $i^\prime$ (blue and red solid lines). The black vertical dashed line corresponds to the best-fit inclination of \cyg. Blue vertical dotted lines correspond to the $1\sigma$ error on the best-fit inclination.}
    \label{fig:inclination_confidence}
\end{figure}

The polarization profiles corresponding to the case of the optically thin co-rotating scatterer in a circular orbit can be decomposed into Fourier series of the orbital longitude $\lambda = 2\pi\phi$ (where $\phi$ is the orbital phase): 
\begin{equation}\label{eq:Fourier}
\begin{aligned}
q_\text{int}  & = q_0 + q_1\cos{\lambda} + q_2\sin{\lambda} + q_3\cos{2\lambda} + q_4\sin{2\lambda}, \\
u_\text{int}  & = u_0 + u_1\cos{\lambda} + u_2\sin{\lambda} + u_3\cos{2\lambda} + u_4\sin{2\lambda}.
\end{aligned}
\end{equation}
We employ Bayesian inference implemented as the Markov chain Monte Carlo (MCMC, \citealt{Goodman2010}) ensemble sampler in the \textit{emcee} package \citep{emcee} in \textsc{python} to fit the orbital profiles of Stokes parameters, observed during season S2 with Eq.~\eqref{eq:Fourier}.
The best-fit curves are shown in Fig.~\ref{fig:set2_fourier_modelling}.
Following the approach described in \citet{Drissen1986} and \citet{Kravtsov2020}, we use the obtained Fourier coefficients to derive the inclination $i$ of the binary   
\begin{equation}\label{eq:Drissen_i}
\left(\frac{1 - \cos{i}}{1 + \cos{i}}\right)^4 = \frac{(u_3 + q_4)^2 + (u_4 - q_3)^2}{(u_4 + q_3)^2 + (u_3 - q_4)^2} , 
\end{equation}
and the position angle $\Omega$ of the orbital axis on the sky 
\begin{equation}\label{eq:Drissen_omega}
\tan{2\Omega} = \frac{A + B}{C + D},\\
\end{equation}
where
\begin{equation}
\begin{aligned}
A &= \frac{u_4 - q_3}{(1 -\cos{i})^2},\quad B = \frac{u_4 + q_3}{(1 +\cos{i})^2} ,\\
C &= \frac{q_4 - u_3}{(1 +\cos{i})^2},\quad D = \frac{u_3 + q_4}{(1 -\cos{i})^2}.
\end{aligned}
\end{equation}

By fitting orbital polarization profiles obtained in S2 with Eq.~\eqref{eq:Fourier}, we got formal values of the inclination $i = 125\degr \pm 5\degr$ ($i > 90\degr$ indicates the clockwise apparent motion of the compact object on the sky) and the position angle $\Omega = 129\degr \pm 5\degr$ of the orbital axis on the sky. However, the formal errors on the estimated orbital parameters obtained from the error propagation are underestimated and hence do not correspond to their actual confidence intervals, which are determined primarily by the internal properties of the model (\ref{eq:Fourier}) and the amplitude of the stochastic variability in the data.
Indeed, the inclination estimates corresponding to the best-fit Fourier coefficients are always biased towards higher values \citep{Aspin1981,Simmons1982,Wolinski1994}. 
The confidence intervals on the orbital parameters for different signal-to-noise ratios can be obtained using Fig.~4 of  \citet{Wolinski1994}: $1\sigma$ and $2\sigma$ confidence intervals are calculated for four levels of data quality given by $\gamma = 0.5N(A / \sigma_p)^2$, where, $\sigma_p$ is the standard deviation of noise in the data, $A$ is an amplitude of polarimetric variability, and $N$ is the number of observations.
Our value of $\gamma_{\rm obs} = 0.5 \times 100 \times (6.7)^{2} \approx 2200$ lies between their grid points ($\gamma = 120000$ and $\gamma = 300$), hence to calculate the confidence intervals on the inclination that we got for our S2 data, we performed our own Monte Carlo simulations following the procedure described in \citet{Wolinski1994}: we modeled the Stokes parameters for different values of $i$ ranging from 90\degr\, to 180\degr\, using the standard \citet{BME} model (eq.~6 in that paper).
Then, we simulated the Gaussian noise in $q$ and $u$ by adding the fluctuations of variance $\sigma_p^2 = 0.5NA^2 / \gamma_{\rm obs}$.
The Fourier model (Eq.~\ref{eq:Fourier}) was then fitted to the simulated data using the MCMC approach and inclination $i^\prime$ was calculated using Eq.~\eqref{eq:Drissen_i}.
In Fig.~\ref{fig:inclination_confidence} we show the inclination estimates $i^\prime$ (black points) with the $1\sigma$ and $2\sigma$ confidence intervals (blue and red solid lines) as a function of the true, input inclination~$i$. 

The inclination $i \approx 125\degr$ that we derived from the best-fit Fourier coefficients of \cyg in $V$ and $R$ bands (shown as a black dashed vertical line in Fig.~\ref{fig:inclination_confidence}) is close to the so-called critical angle $i^\prime_\text{crit}$ -- the angle above which the $1\sigma$ confidence interval on the orbital inclination extends to $i = 180\degr$ \citep{Wolinski1994}. That means that using high-precision polarimetry, we can only put a lower limit of $180\degr > i > 120\degr$ on the value of inclination of \cyg orbit. We note that previous polarimetrically derived inclination values \citep{Kemp1978, Dolan1989, Nagae2009} are most likely overestimated because they were obtained by modeling the data with larger error bars, for which the critical angle is expected to be smaller than  $i^\prime_\text{crit} \sim 130\degr$.
Our lower limit on the inclination $i > 120\degr$ and the clockwise direction of the orbital motion on the sky are consistent with the value $i = 153\degr \pm 1\degr$ from \citet{Miller-Jones2021}. 

In contrast to the inclination, the value $\Omega$, obtained from Eq.~\eqref{eq:Drissen_omega} is an unbiased estimate of the true position angle of the projection of the orbital axis on the sky. We use the same MCMC approach as for inclination to calculate the confidence interval on this angle. Our value $\Omega = 129\degr \pm 10\degr$ (or $\Omega = 129\degr - 180\degr = - 51\degr \pm 10\degr$ because of the $\pm180\degr$ ambiguity) is consistent within $3\sigma$ with those, determined by the direction of the intrinsic polarization $\theta_\mathrm{int} \approx  -25\degr$ and the position angle of the jet on the sky $\Omega_\mathrm{jet} \approx -26\degr$ \citep{Miller-Jones2021}.    

We emphasize that this low accuracy is not a result of the polarization measurement errors (which are smaller than $0.01\%$ for the whole set of our data). Fig.~\ref{fig:set2_fourier_modelling} shows a remarkable intrinsic scatter of S2 data points around the fit curves, especially noticeable for the Stokes $q$ parameter. This aperiodic noise is explained by the presence of an additional sub-orbital variability component which appears on the time scales shorter than one orbital cycle. Thus, the key assumption on co-rotation of the light scattering material over (at least) a few consecutive orbital cycles does not hold for \cyg binary system. Therefore, the `traditional' Fourier fit up to second harmonics made to polarization data cannot provide meaningful estimates of orbital inclination regardless of data quality, quantity, and sampling frequency.

\subsection{Eccentric model}\label{sect:eccentric_model}

While for a circular orbit theory predicts a smooth change in polarization with the dominant second harmonic of the orbital period, the eccentricity of the orbit shifts all the changes in the polarization towards the periastron. Indeed, the polarization depends on the scattering angle, which changes according to the orbital motion of the scattering cloud. In the case of eccentric orbit, this angle changes with the different rates in different parts of the orbit, resulting in unequal distances between consecutive maxima/minima of orbital Stokes parameters curves \citep[this effect was observed for the binary with $e\approx 0.4$ in][]{Berdyugin1998}. Therefore, the orbital curves of Stokes parameters can be used for independent estimation of the orbital eccentricity.
We adopted the Thomson scattering model from the appendix of \cite{Kravtsov2020} to describe the orbital changes in the polarization of \cyg. 
By fitting this model to the \textit{V}-band Season 2 data, we were able to put $3\sigma$ upper limit on the eccentricity of \cyg orbit to be $e < 0.08$.

\begin{figure}
\centering
\includegraphics[width=\linewidth]{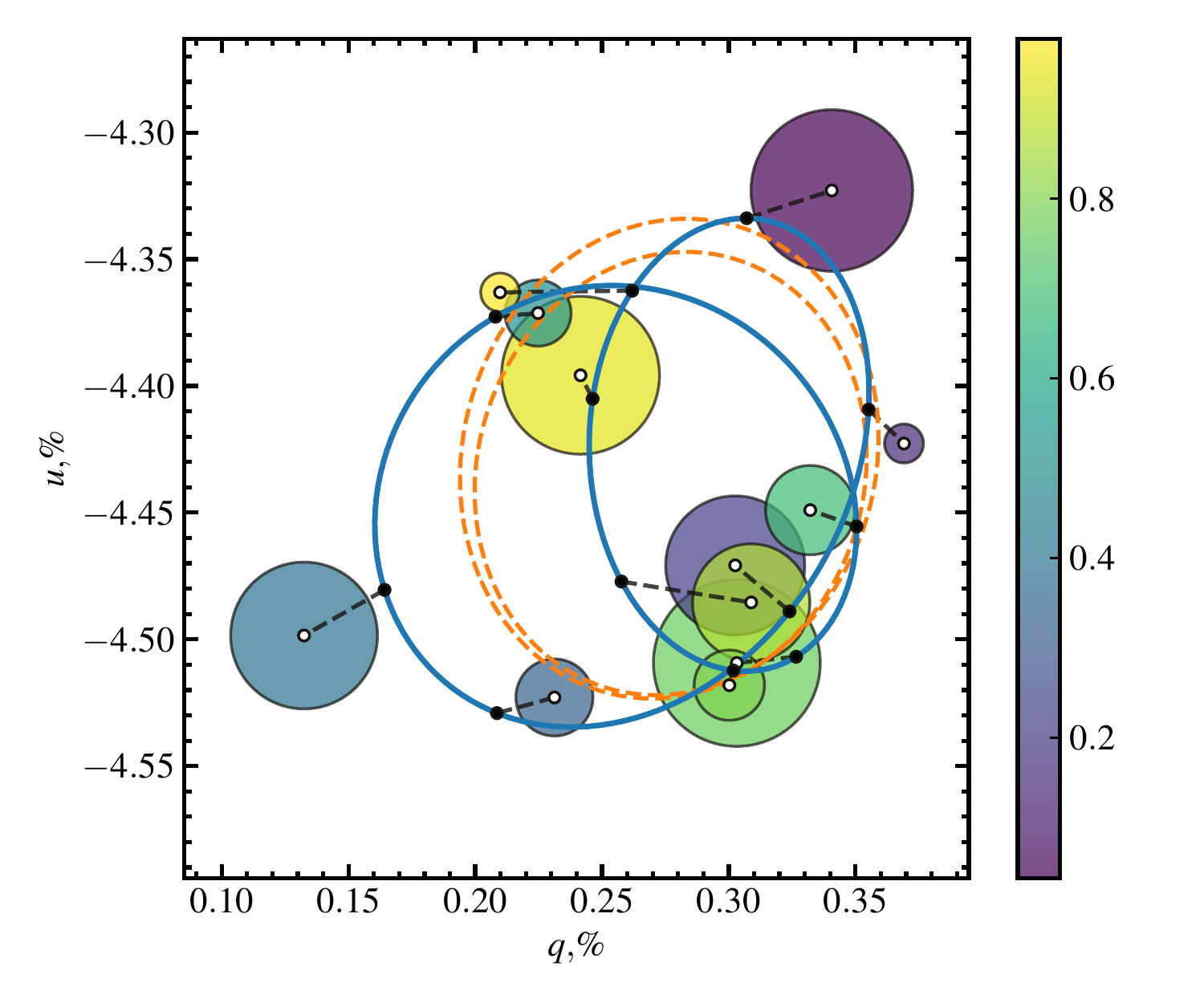}
\caption{Average Stokes parameters of \cyg, obtained during Season~2 in the \textit{R} band. Color coding and the size of the circle correspond to the orbital phase and $1\sigma$ uncertainty, respectively. The blue solid curve correspond to the best fit with the Fourier series given by Eq.~\eqref{eq:Fourier}. The orange dashed line corresponds to the to the best-fit model of a scattering cloud on an eccentric orbit from the Appendix of \citet{Kravtsov2020}.}
\label{fig:quplane_models}
\end{figure}

Fig.~\ref{fig:quplane_models} shows the ($q$, $u$)-plane of average Stokes parameters of \cyg obtained during Season 2 together with the best-fits with the Fourier series (Eq.~\ref{eq:Fourier}) and the model of Thomson scattering by a cloud on an eccentric orbit \citep{Kravtsov2020}. We see that the latter model (being a special case of the general BME model for a symmetrically distributed matter about the orbital plane) cannot reproduce the pretzel-like shape of the trace left by orbital variations of \cyg on the ($q$, $u$)-plane -- the additional source of asymmetry is needed. To explain a similar pattern, \cite{Kemp1978} proposed the model in which the scattering region is eclipsed by the secondary body for half the orbit. This model requires a high ($90\degr > i > 130\degr$) orbital inclination, which contradicts the latest results (including this article). 

\subsection{Polarization by Thomson scattering off a precessing accretion disk}
\label{sec:prec_disk}
\begin{figure}
\centering
\includegraphics[width=\linewidth, trim={3cm 0 3cm 0}]{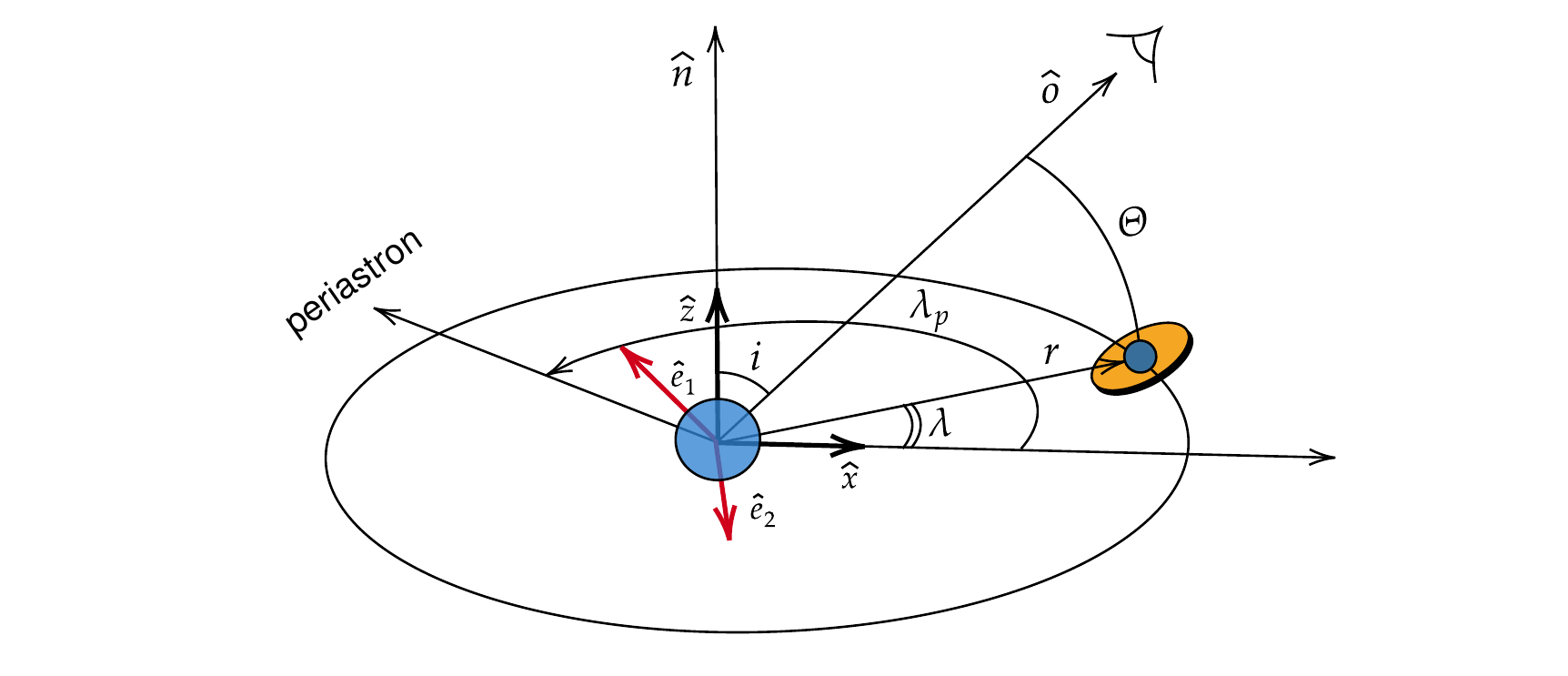}
\caption{\label{fig:orbit_geometry} 
Geometry of the system.} 
\end{figure}

\begin{figure}
\centering
\includegraphics[width=\linewidth, trim={0 0 3cm 0}]{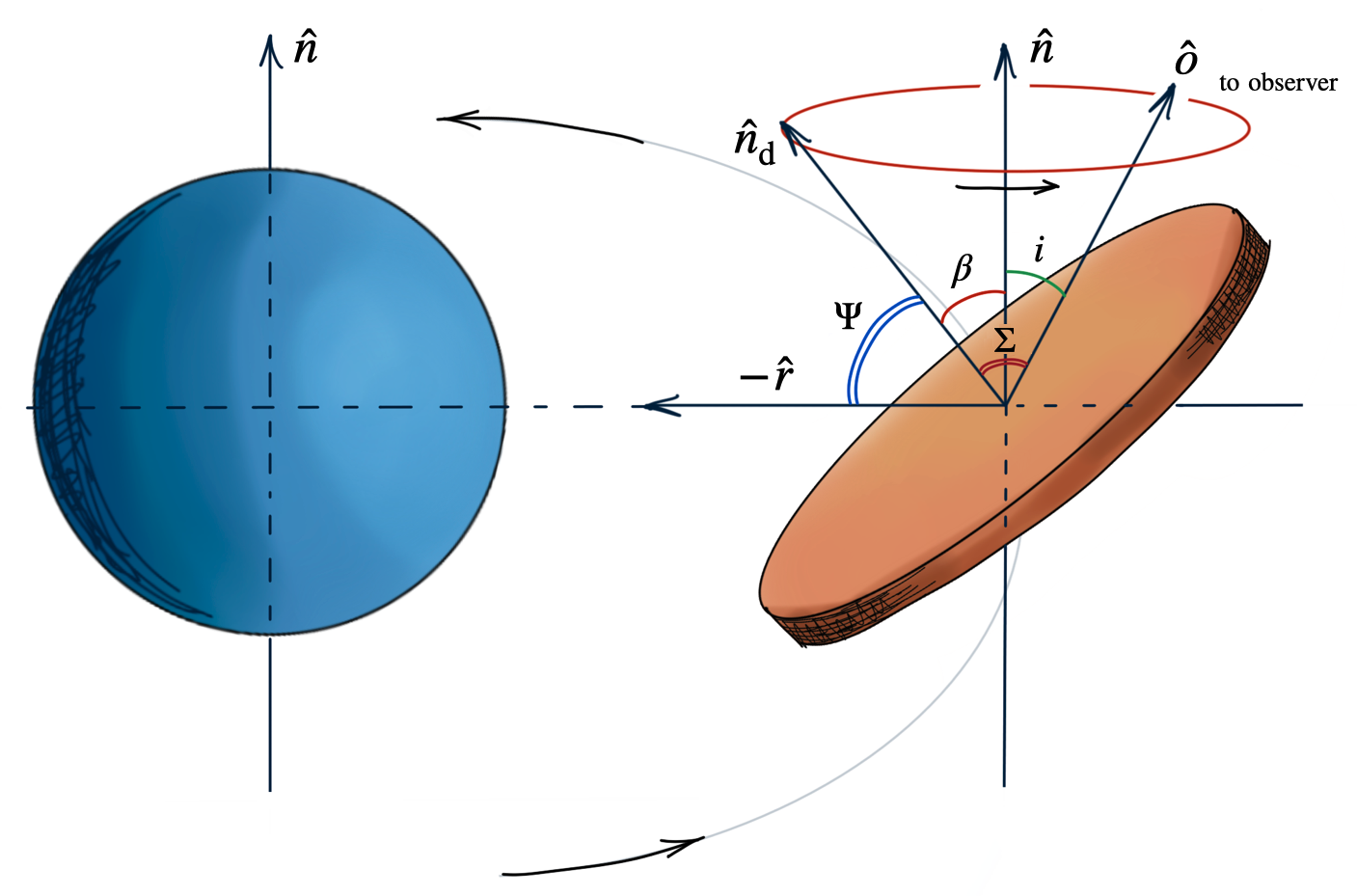}
\caption{\label{fig:model_geometry} 
Geometry of the precessing disk. }
\end{figure}

In this section, we present a model of polarization from Thomson scattering by a tilted, precessing accretion disk, which can naturally explain the asymmetric pattern of the polarization variability observed in \cyg  without requiring a highly inclined or eccentric orbit.
We consider the following geometry: the orbit with the eccentricity $e$ is inclined by angle $i$ to the line of sight \unit{o} (Fig.~\ref{fig:orbit_geometry}). 
The accretion disk, surrounded by a cloud of electrons, rotates around the optical companion together with the compact object.
The disk axis $\unit{n}_{\rm d}$ is inclined by angle $\beta$ to the orbital axis $\unit{n}$ (see Fig.~\ref{fig:model_geometry}). 
The axis of the disk can precess about the orbital axis with the period $T_\mathrm{sup}$.

To describe the orbital motion, we introduce the coordinate system ($\unit{x}$, $\unit{y}$, $\unit{z}$), in which $\unit{z}$-axis is directed along the orbital axis $\unit{n}$, vector $\unit{x}$ lies in the orbital plane and its projection on the sky is directed to the south, and vector $\unit{y}$ forms the right-handed basis. 
In this basis, $\unit{o} = (\sin i, 0, \cos i)$, $\unit{n} = (0, 0, 1)$, and $\unit{n}_{\rm d} = (\sin\beta \cos\gamma,\ \sin\beta \sin\gamma,\ \cos\beta)$.
The angle $\gamma$ is the azimuth of the projection of the disk axis onto the orbital plane  measured from $\unit{x}$ to $\unit{y}$.
To describe the polarization, we use the polarization basis ($\unit{e}_1$, $\unit{e}_2$), in which the vector $\unit{e}_1 = (-\cos i, 0, \sin i)$ lies along the projection of the vector \unit{n} on the plane of the sky, and $\unit{e}_2 = (0, -1, 0)$ is perpendicular to $\unit{e}_1$ and lies in the plane of the orbit.

The distance between the compact object and optical companion varies with the orbital longitude $\lambda$, measured from $\unit{x}$ to $\unit{y}$ and can be expressed as
\begin{equation} 
r(\lambda)= \frac{a(1-e^2)}{1+ e \cos (\lambda-\lambda_{\rm p})}, 
 \end{equation} 
where $a$ is the semi-major axis of the orbit and $\lambda_{\rm p}$ is the longitude of the periastron. The unit vector pointing towards the compact object is
\begin{equation} 
\unit{r}= (\cos\lambda, \sin\lambda, 0)  
 ,\end{equation}
and the scattering angle $\Theta$ is given by 
\begin{equation} 
\mu= \cos\Theta = \unit{r} \cdot   \unit{o}  =  \sin i \cos \lambda . 
 \end{equation}

The observed flux  $F_{\rm tot} = F_* + F_{\rm sc}$ is the sum of the flux produced by the optical companion $F_*$ and the scattered flux $F_{\rm sc}$. We assume that the latter is produced by Thomson scattering (in an optically thin regime) of stellar radiation by the accretion disk and surrounding cloud of electrons. In that case, the angular distribution of scattered luminosity can be represented as 
\begin{equation} 
L_{\rm sc}(\mu)  = L_* \ f_{\rm sc}\  l(\mu) =  \frac{3}{8} (1+\mu^2)\ L_*\ (f^{\rm cloud}_{\rm sc} + f^{\rm disk}_{\rm sc}), 
 \end{equation} 
where $l(\mu) = 3(1+\mu^2)/8$ is the Thomson scattering indicatrix, $f^{\rm cloud}_{\rm sc}$ and $f^{\rm disk}_{\rm sc}$  are fractions of radiation scattered by the cloud and the disk respectively. In both cases, this fraction is proportional to the total number of free electrons $N_{\rm e}$ in a cloud/disk and drops with the distance as $1/r^2(\lambda)$:
\begin{equation} 
f^{\rm cloud}_{\rm sc} = \frac{N^{\rm cloud}_{\rm e}\sigma_{\rm T}}{4\pi r^2(\lambda)},
\end{equation}
\begin{equation} 
f^{\rm disk}_{\rm sc} = \frac{N^{\rm disk}_{\rm e}\sigma_{\rm T}}{4\pi r^2(\lambda)}\cos{\Psi},
\end{equation} 
where $\cos{\Psi} = -\unit{n}_{\rm d} \cdot \unit{r} = - \sin{\beta}\cos{\gamma}\cos{\lambda} - \sin{\beta}\sin{\gamma}\sin{\lambda} = - \sin\beta\cos{(\gamma - \lambda)}$.
The $\cos{\Psi}$ term is proportional to the effective area of the disk intercepting the stellar radiation, which depends on the position of the disk on the orbit $\lambda$ and the orientation of its axis $\unit{n}_{\rm d}$, defined by two angles: the inclination $\beta$ of disk and its azimuth $\gamma$. The latter angle can change with time due to precession as $\gamma = \pm 2\pi\varphi_{\rm sup} + \gamma_0$, where $\varphi_{\rm sup}$ is the precession phase and $\gamma_0$ is the angle $\gamma$ at zero phase. 
The sign determines the direction (counterclockwise of clockwise) of the precession.

We scale $f^{\rm cloud}_{\rm sc}$ and $f^{\rm disk}_{\rm sc}$ to the typical values  $f^{\rm cloud}_0$ and $f^{\rm disk}_0$ as
\begin{equation} 
f^{\rm cloud}_{\rm sc}  = f^{\rm cloud}_0\ \left[\frac{a(1-e^2)}{r(\lambda)}\right]^2 = f^{\rm cloud}_0 \ [1+ e \cos (\lambda-\lambda_{\rm p})]^2, 
 \end{equation} 
\begin{equation} 
f^{\rm disk}_{\rm sc}  = f^{\rm disk}_0\ \left[\frac{a(1-e^2)}{r(\lambda)}\right]^2\cos{\Psi}= f^{\rm disk}_0 \ [1+ e \cos (\lambda-\lambda_{\rm p})]^2\cos{\Psi}.
\end{equation}

The PD of scattered radiation in Thomson regime can be expressed in terms of the scattering angle $\Theta$ as 
\begin{equation} 
P_{\rm sc}= \frac{1-\mu^2}{1+\mu^2} = \frac{\sin^2\Theta}{1+\cos^2\Theta}. 
 \end{equation} 
The observer measures the PD $P= F_{\rm sc} P_{\rm sc} / F_{\rm tot}$ of the total flux $F_{\rm tot}$, most of which is the unpolarized and produced by the optical companion star. The polarized flux of the scattered radiation is 
\begin{equation} 
F_{\rm sc} P_{\rm sc} = F_*\  f_{\rm sc}\ l(\mu)\   \frac{1-\mu^2}{1+\mu^2}  
= F_*\  f_{\rm sc}\   \frac{3}{8}\ (1-\mu^2) .
 \end{equation} 
Therefore, the total PD 
\begin{equation} \label{eq:pol_tot}
P = \frac{F_{\rm sc} P_{\rm sc}}{F_{\rm tot}}  \approx \frac{3}{8}\ \left[f^{\rm cloud}_{\rm sc} + f^{\rm disk}_{\rm sc}\right] \ (1-\mu^2), 
\end{equation} 
where we assumed that $F_{\rm sc} \ll F_{*}$ and substituted $f_{\rm sc} = f^{\rm cloud}_{\rm sc} + f^{\rm disk}_{\rm sc}$. 

The normalized Stokes parameters of linear polarization are defined as $q=P \cos(2\rchi)$ and $u=P\sin(2\rchi)$, where $\rchi$ is the position angle of the polarization pseudo-vector $\unit{p}$ in the polarization basis ($\unit{e}_1$, $\unit{e}_2$):
\begin{equation} 
\unit{p} =  \frac{\unit{o} \times  \unit{r}}{|\unit{o} \times  \unit{r}|} 
= \frac{1}{\sin\Theta}\left(-\cos i \sin\lambda,  \cos i \cos\lambda, \sin i \sin\lambda \right),
\end{equation} 
where $\sin\Theta= \sqrt{1-\mu^2}= \sqrt{1-\sin^2 i \cos^2\lambda}$.
Thus, the expressions for the Stokes parameters can be written as 
\begin{equation} \label{eq:stokes_expr}
\begin{aligned}
q &=  \frac{3}{8} \left[ f^{\rm cloud}_{\rm sc} + f^{\rm disk}_{\rm sc} \right]\ (1-\mu^2) \cos(2\rchi), \\
u &=  \frac{3}{8} \left[ f^{\rm cloud}_{\rm sc} + f^{\rm disk}_{\rm sc} \right]\ (1-\mu^2) \sin(2\rchi),
\end{aligned}
\end{equation}
where the polarization angle $\rchi$ is defined by the expressions: 
\begin{eqnarray} 
\cos\rchi &=& \unit{e}_1 \cdot  \unit{p} = \frac{\sin\lambda}{\sin\Theta}, \\
\sin\rchi &=& \unit{e}_2 \cdot  \unit{p} =  - \frac{\cos i \cos\lambda}{\sin\Theta}.
\end{eqnarray} 
The explicit expressions for $\cos(2\rchi)$ and $\sin(2\rchi)$ are:
\begin{eqnarray} \label{eq:double_chi_1}
\cos(2\rchi) &=& 
\frac{\sin^2 i - (1+\cos^2 i) \cos(2\lambda)}{2\sin^2\Theta}, \\
\sin(2\rchi) &=& - \frac{\cos i  \sin(2\lambda)}{\sin^2\Theta}.  \label{eq:double_chi_2}
\end{eqnarray} 
Combining (\ref{eq:stokes_expr}) with (\ref{eq:double_chi_1}), (\ref{eq:double_chi_2}), we get
\begin{equation}\label{eq:model_Stokes}
\begin{aligned}
q &=  \frac{3}{16}  \left[\sin^2 i - \left(1\!+\!\cos^2\! i\right) \cos2\lambda\right]\left[ f^{\rm cloud}_{\rm sc} + f^{\rm disk}_{\rm sc} \right] , \\
u &= -\frac{3}{8} \cos i\  \sin2\lambda \left[ f^{\rm cloud}_{\rm sc} + f^{\rm disk}_{\rm sc} \right].
\end{aligned}
\end{equation}

The $f^{\rm disk}_{\rm sc}$ term depends on $\cos{\Psi}$,
reflecting the difference in the amount of scattered radiation for different orientations of the disk axis $\unit{n}_{\rm d}$ relative to the source of the light. In our model, the disk is not transparent: it has two sides (top and bottom), only one of which is illuminated at any given time. The top of the disk is illuminated when $\cos{\Psi} > 0$, the bottom of the disk is bright when $\cos{\Psi} < 0$. The top of the disk is visible for the observer when $\cos{\Sigma}=\unit{n}_{\rm d} \cdot \unit{o} > 0$ and the bottom of the disk is visible when $\cos{\Sigma} < 0$. Therefore, the disk is illuminated and visible only when the product $\cos{\Psi}\cos{\Sigma}$ is positive, 
or $f^{\rm disk}_{\rm sc} = 0$ when $\cos{\Psi}\cos{\Sigma} \leq 0$.

To compare the calculations with the observed Stokes parameters $(q_{\rm obs}$, $u_{\rm obs})$, one needs to take into account the orientation of the orbit on the sky: the projection of the orbital axis on the sky makes angle $\Omega$ with direction to the north  (we note that $\Omega$ defined this way differs by $\pi/2$ from the longitude of the ascending node commonly used instead). 
The observed Stokes parameters $(q_{\rm obs}, u_{\rm obs})$ can be obtained by rotating vector $(q,u)$ by angle $2\Omega$: 
\begin{eqnarray} \label{eq:q_rotate}
q_\text{obs} &=&  q\cos(2\Omega)-u\sin(2\Omega),\\
 \label{eq:u_rotate}
u_\text{obs} &=&  q\sin(2\Omega)+u\cos(2\Omega).
\end{eqnarray}

\begin{figure}
\centering
\includegraphics[width=\linewidth]{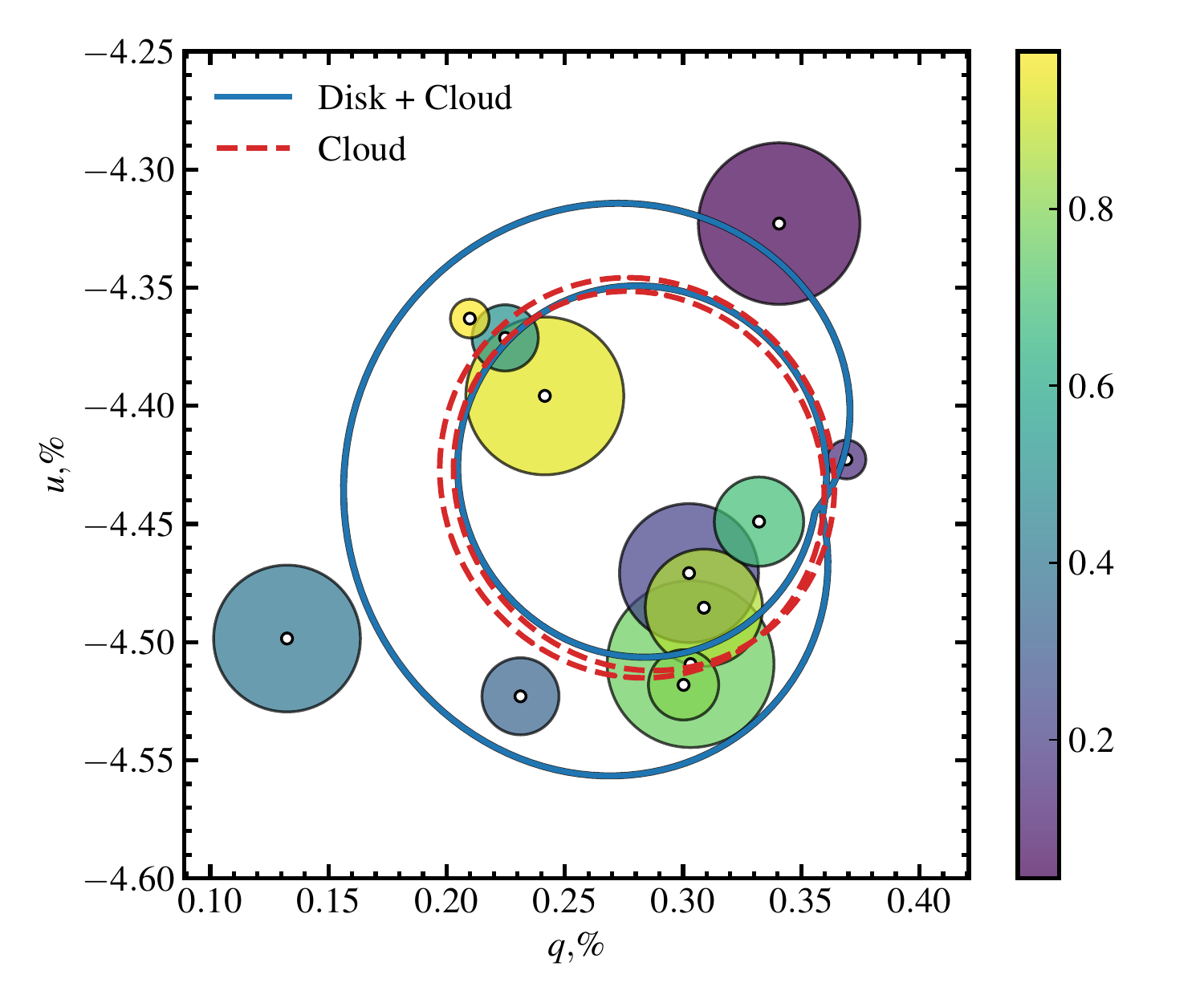}
\caption{\label{fig:modeled_quplane}
Same as Fig.~\ref{fig:quplane_models}, but showing the model curves at the ($q$, $u$)-plane calculated with (blue solid line) and without (red dashed line) accounting for scattering by the accretion disk.}
\end{figure}

\begin{figure}
\includegraphics[width=0.9\linewidth]{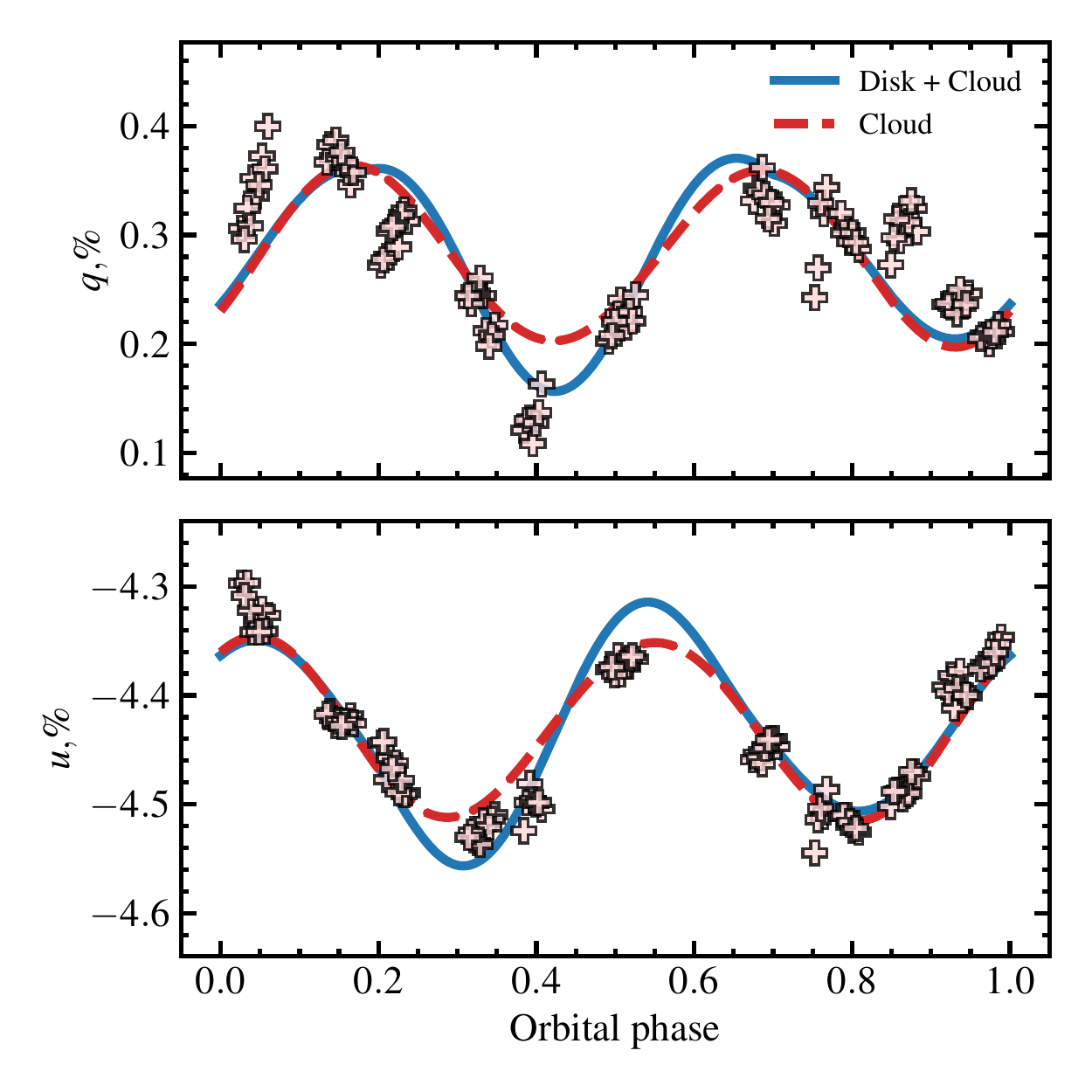}
\caption{\label{fig:modeled_q_and_u}
Stokes parameters of \cyg, obtained during Season 2 (light crosses) together with the best-fit models with (blue solid line) and without (red dashed line) taking into account scattering by the tilted accretion disk.}
\end{figure}

The modeled Stokes parameters are functions of the orbital longitude $\lambda$ and need to be computed as functions of the orbital phase $\phi$. While for the circular (or nearly circular) orbit $\lambda$ can be calculated as $\lambda = 2\pi(\phi + \phi_{\rm p}) + \lambda_{\rm p}$, where $\phi_{\rm p}$ is the phase of the periastron, for the eccentric orbit one needs to solve the Kepler's equation: from the true anomaly of the orbit $\lambda- \lambda_\text{p}$, we can find the eccentric anomaly $E$ 
\begin{equation} \label{eq:eccent_anom}
\tan\left(\frac{E}{2}\right)=\sqrt{\frac{1-e}{1+e}} \tan\left(\frac{\lambda- \lambda_\text{p}}{2}\right) ,
\end{equation} 
and then the mean anomaly $M$
\begin{equation} \label{eq:meananom}
M=E-e\sin E   ,
\end{equation} 
which can be converted to the orbital phase as    
\begin{equation} \label{eq:phiorb}
\phi_\text{orb} = M/(2\pi) + \phi_\text{p} .
\end{equation}

Thus, the free parameters of the model are the inclination $i$, eccentricity $e$, longitude of the periastron $\lambda_{\rm p}$, the position angle of the projection of the orbit axis on the sky $\Omega$, the phase of the periastron $\phi_{\rm p}$, the inclination of the disk $\beta$ and its initial position angle $\gamma_0$ in the orbital plane, the period of precession $T_{\rm sup}$ (which can be set to infinite for non-precessing case), the scattering fractions $f^{\rm disk}_0$ and $f^{\rm cloud}_0$, and constant Stokes parameters $q_0$ and $u_0$, which describe average polarization.

We fit the described model to S2 data, adopting the orbital parameters of \cyg:  eccentricity $e=0.02$, inclination $i=153\degr$ \citep{Miller-Jones2021}, inclination of the disk $\beta = 20\degr$ \citep{Ibragimov2007}. 
The position angle of the orbital axis was set to be $\Omega = -26\degr$ to match the position angle of the jet. 
The contributions from the disk and the cloud were assumed to be  $f^{\rm disk}_0 = 0.75$ and $f^{\rm cloud}_0 = 0.25$. 
Since S2 data covers only one full orbital cycle, the precession period $T_{\rm sup}$ was set to be much larger than the orbital period not to  take possible precession of the disk into account. 

The solid blue and dashed red lines in Figs.~\ref{fig:modeled_quplane} and \ref{fig:modeled_q_and_u} show the fits of the model described above to S2 data with and without accounting for scattering off the tilted accretion disk, respectively. 
Although the reduced $\chi^2$ of the fits does not differ dramatically ($\chi^2\left[\rm disk + cloud\right] = 1.01$ versus $\chi^2\left[\rm cloud\right] = 1.23$), the asymmetric, pretzel-like trace of the polarization on the $(q, u)$-plane cannot be reproduced by the scattering cloud alone. 
Additionally, the model with the tilted accretion disk predicts the changes in the shape of the orbital polarization profiles with the precession phase -- if the superorbital variability observed from the radio to the X-rays is related to the disk precession, the pretzel will make a complete turn around its center once per superorbital period $T_{\rm sup}$. 
To detect this effect, a significant part of the superorbital period must be covered with continuous high-precision optical polarimetric observations. 

\section{Summary}
\label{sec:sum}

In this paper, we presented new high-precision polarimetric observations of the BH X-ray binary \cyg.
Combining them with the 12-year-long PMO observations performed in 1975--1987, we were able to study the polarization behavior at the timescales ranging from hours to decades.
The interstellar polarization, which dominates the observed optical polarization $P_{\rm obs} \sim 4.5\%$, was accurately measured and subtracted from the data, allowing us to determine the intrinsic polarization of \cyg. 
The alignment of the X-ray and optical PA, as well as the stability of this angle during the secular PD change, indirectly support our estimate of intrinsic polarization of \cyg. 
Round-the-clock monitoring of the polarization with two telescopes located in different hemispheres allowed us to track the evolution of the polarization within one orbital cycle with a previously unachievable temporal resolution. 
The intrinsic polarization of \cyg shows the orbital variations with two pronounced peaks in the quadratures and two minima in conjugations, most probably produced by Thomson scattering of the companion star radiation by matter gravitationally bound to the black hole. 
The amplitudes of the two consecutive polarization minima measured within one orbital cycle differ significantly, which implies the asymmetry of the scattering matter about the orbital plane.
We suggest that a tilted accretion disk could be the source of such asymmetry. 
We find the misalignment of $\beta \sim 15\degr-30\degr$ to be sufficient to reproduce the orbital behavior of the polarization. 
We modeled orbital variations of the Stokes parameters and constrained the eccentricity $e < 0.08$ and inclination of the orbit $i > 120\degr$.

In addition to the orbital variations, we found a significant change ($\Delta P_\text{int} \approx -0.4\%$) in the average intrinsic PD of \cyg on the timescales of several decades while preserving the constant intrinsic PA.
The decrease of the PD indicates the change in the fraction of scattered radiation, that in turn depends on the amount of scattering material and its effective scattering cross-section. 
This may reflect secular changes in the size/shape of the accretion disk or/and changes in its spatial orientation. 
We note that the asymmetry of the $(q,u)$-plane trace of the polarization can be purely artificial. 
It may result from a complex superposition of periodic, non-periodic variations, and the orbital phase sampling. 
A long-term high-precision monitoring program with good orbital and superorbital coverage is needed to exclude it. 

Analyzing high-precision TESS photometric data, we found the presence of stochastic variations of the flux on the timescales shorter than the orbital period. Together with the stochastic variability found in the optical polarization, this suggests that one or several additional components are at play: pulsations of the optical companion, the presence of the spots on its surface, the wind clumpiness, the eclipses of the bright part of the accretion disk by the infalling matter, and the precession of the accretion disk. 

\section*{Acknowledgements}

This paper is based on observations made with the Nordic Optical Telescope, owned in collaboration by the University of Turku and Aarhus University, and operated jointly by Aarhus University, the University of Turku, and the University of Oslo, representing Denmark, Finland, and Norway, the University of Iceland and Stockholm University at the Observatorio del Roque de los Muchachos, La Palma, Spain, of the Instituto de Astrofisica de Canarias.

The DIPol-2 and DIPol-UF polarimeters were built in cooperation between the University of Turku, Finland, and the Leibniz-Institut f\"{u}r Sonnenphysik, Germany.
We are grateful to the Institute for Astronomy, University of Hawaii, for the allocated observing time. 
VK acknowledges support from the Vilho, Yrj\"o and Kalle V\"ais\"al\"a Foundation and  Suomen Kulttuurirahasto. AAZ has been supported by the Polish National Science Center under the grant 2019/35/B/ST9/03944.

\bibliographystyle{aa}
\bibliography{aanda.bib}


\end{document}